%% file: emulator.tex
\documentclass[final,5p,times,twocolumn]{elsarticle}

\input{Content/preamble}

\begin{document}

\input{Content/frontmatter}
\input{Content/introduction}
\input{Content/model}
\input{Content/procedures}
\input{Content/gradients}
\input{Content/weaklensing}
\input{Content/software}
\input{Content/results}
\input{Content/conclusions}
\input{Content/acknowledgement}
\input{Content/code}

\bibliographystyle{elsarticle-harv}\biboptions{semicolon,round,sort,authoryear}
\bibliography{references}

\end{document}

%% file: Content/preamble.tex
\usepackage{graphicx}	
\usepackage{amsmath}	
\usepackage[nice]{nicefrac}
\usepackage{amssymb}	
\usepackage{bm}
\usepackage{fontawesome}

\usepackage{hyperref}
\usepackage{graphicx}
\usepackage{xcolor}
\usepackage{multirow}
\usepackage{float}
\usepackage{booktabs}
\usepackage{subfig}
\usepackage{lineno}
\usepackage{seqsplit}
\usepackage{verbatim}
\usepackage{ulem}
\usepackage{url}
\usepackage{tabularx}
\usepackage{listings}

\definecolor{codegreen}{rgb}{0,0.6,0}
\definecolor{codegray}{rgb}{0.5,0.5,0.5}
\definecolor{codepurple}{rgb}{0.58,0,0.82}
\definecolor{backcolour}{rgb}{0.95,0.95,0.92}

\lstdefinestyle{mystyle}{
    backgroundcolor=\color{backcolour},   
    commentstyle=\color{codegreen},
    keywordstyle=\color{magenta},
    numberstyle=\tiny\color{codegray},
    stringstyle=\color{codepurple},
    basicstyle=\ttfamily\footnotesize,
    breakatwhitespace=false,         
    breaklines=true,                 
    captionpos=b,                    
    keepspaces=true,                 
    numbers=left,                    
    numbersep=5pt,                  
    showspaces=false,                
    showstringspaces=false,
    showtabs=false,                  
    tabsize=2
}

\usepackage{color}
\definecolor{darkblue}{rgb}{0.,0.,0.4}
\definecolor{darkred}{rgb}{0.5,0.,0.}
 \hypersetup{
     colorlinks=true,
     linkcolor=blue,
     filecolor=red,
     citecolor=red,      
     urlcolor=cyan,
     }

\definecolor{ForestGreen}{RGB}{34, 139, 34}

\newcommand{\bb}[1]{\mathbb{#1}}
\newcommand{\bs}[1]{\boldsymbol{#1}}
\newcommand{\mc}[1]{\mathcal{#1}}
\newcommand{\tm}[1]{\textrm{#1}}
\newcommand{\sans}[1]{\bm{\mathsf{#1}}}


%% file: Content/frontmatter.tex
\begin{frontmatter}

\title{Kernel-Based Emulator for the 3D Matter Power Spectrum from CLASS}

\author[inst1]{Arrykrishna Mootoovaloo}
\author[inst1]{Andrew H. Jaffe}
\author[inst1]{Alan F. Heavens}
\author[inst1]{Florent Leclercq}

\affiliation[inst1]{organization={Imperial Centre for Inference and Cosmology (ICIC),  Department of Physics, Blackett Laboratory, Imperial College},
            addressline={Prince Consort Road}, 
            city={London},
            postcode={SW7 2AZ}, 
            country={UK}}

\begin{abstract}
The 3D matter power spectrum, $P_{\delta}(k,z)$ is a fundamental quantity in the analysis of cosmological data such as large-scale structure, 21cm observations, and weak lensing. Existing computer models (Boltzmann codes) such as CLASS can provide it at the expense of immoderate computational cost. In this paper, we propose a fast Bayesian method to generate the 3D matter power spectrum, for a given set of wavenumbers, $k$ and redshifts, $z$. Our code allows one to calculate the following quantities: the linear matter power spectrum at a given redshift (the default is set to 0); the non-linear 3D matter power spectrum with/without baryon feedback; the weak lensing power spectrum. The gradient of the 3D matter power spectrum with respect to the input cosmological parameters is also returned and this is useful for Hamiltonian Monte Carlo samplers. The derivatives are also useful for Fisher matrix calculations. In our application, the emulator is accurate when evaluated at a set of cosmological parameters, drawn from the prior, with the fractional uncertainty, $\Delta P_{\delta}/P_{\delta}$ centred on 0. It is also $\sim 300$ times faster compared to CLASS, hence making the emulator amenable to sampling cosmological and nuisance parameters in a Monte Carlo routine. In addition, once the 3D matter power spectrum is calculated, it can be used with a specific redshift distribution, $n(z)$ to calculate the weak lensing and intrinsic alignment power spectra, which can then be used to derive constraints on cosmological parameters in a weak lensing data analysis problem. The software ($\texttt{emuPK}$) can be trained with any set of points and is distributed on Github, and comes with a pre-trained set of Gaussian Process (GP) models, based on 1000 Latin Hypercube (LH) samples, which follow roughly the current priors for current weak lensing analyses.
\end{abstract}

\begin{keyword}

Kernel \sep Gaussian Process \sep Emulation \sep Large Scale Structures

\end{keyword}

\end{frontmatter}

%% file: Content/introduction.tex
\section{Introduction}
\label{sec:introduction}

The 3D matter power spectrum, $P_{\delta}(k,z)$ is a key quantity which underpins most cosmological data analysis, such as galaxy clustering, weak lensing, 21 cm cosmology and various others. Crucially, the calculation of other (derived) power spectra can be fast if $P_{\delta}(k,z)$ is precomputed. In practice, the latter is the most expensive component and can be calculated either using Boltzmann solvers such as CLASS or CAMB, or via simulations, which can be computationally expensive depending on the resolution of the experiments.

For the past 30 decades or so, with the advent of better computational facilities, various techniques have been progressively devised and applied to deal with inference in cosmology. In brief, some of these techniques include Monte Carlo (MC) sampling, variational inference, Laplace approximation and recently we are witnessing other new approaches such as density estimation \citep{2018MNRAS.477.2874A, 2019MNRAS.488.4440A, 2019MNRAS.488.5093A} which makes use of tools like Expectation-Maximisation (EM) algorithm and neural networks (NN). Recently, \cite{2018PhRvD..97h3004C} designed the information maximising neural networks (IMNNs) to learn nonlinear functions of data that maximise Fisher information. In this paper, we explore another branch of Machine Learning (ML) which deals with kernel techniques.    

\begin{figure*}
    \centering
    \subfloat[3D Matter Power Spectrum, $P_{\delta}(k,z)$]{{\includegraphics[width=0.45\textwidth]{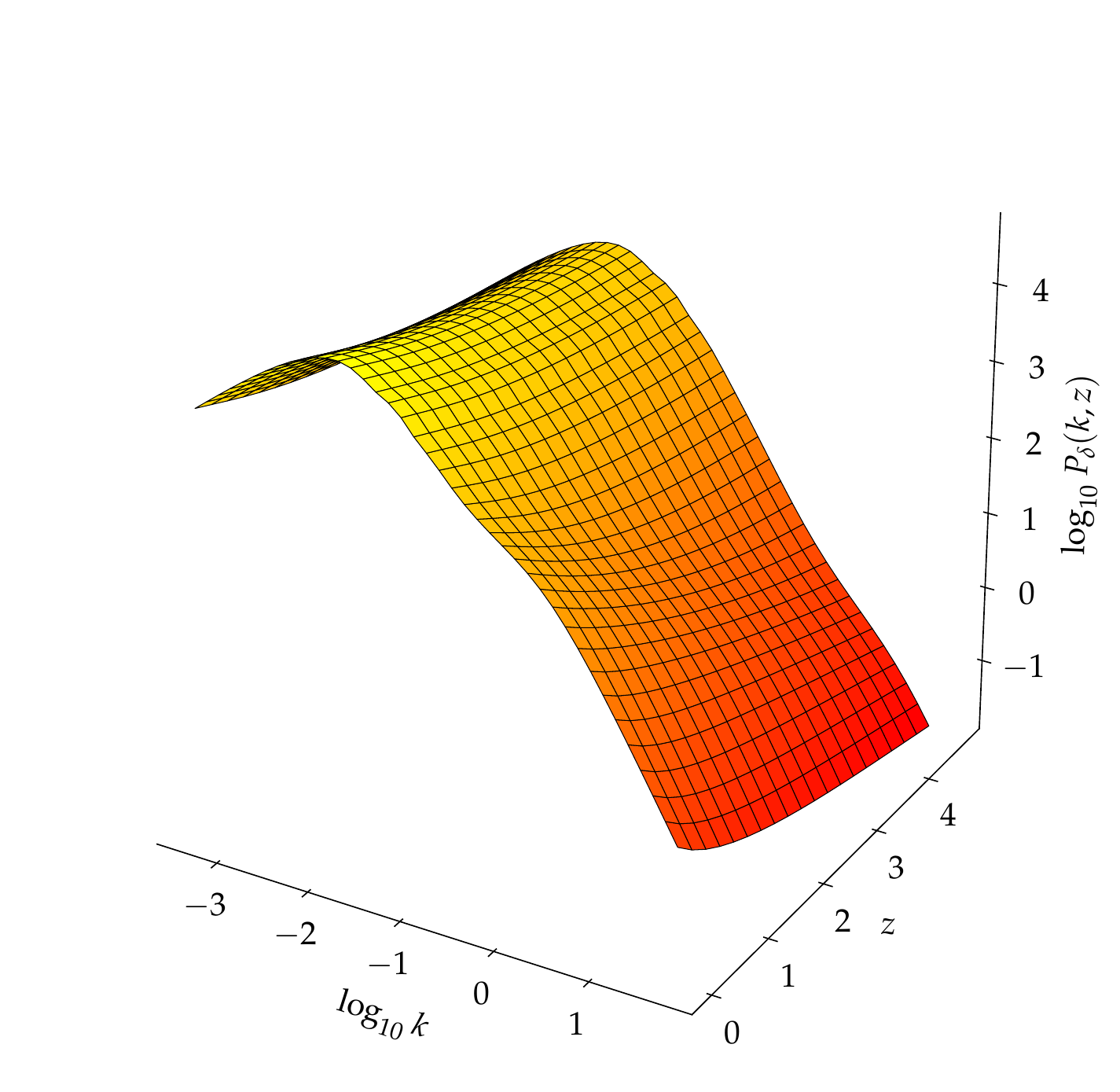}}}
    \qquad
    \subfloat[The function $1+q(k,z)$]{{\includegraphics[width=0.45\textwidth]{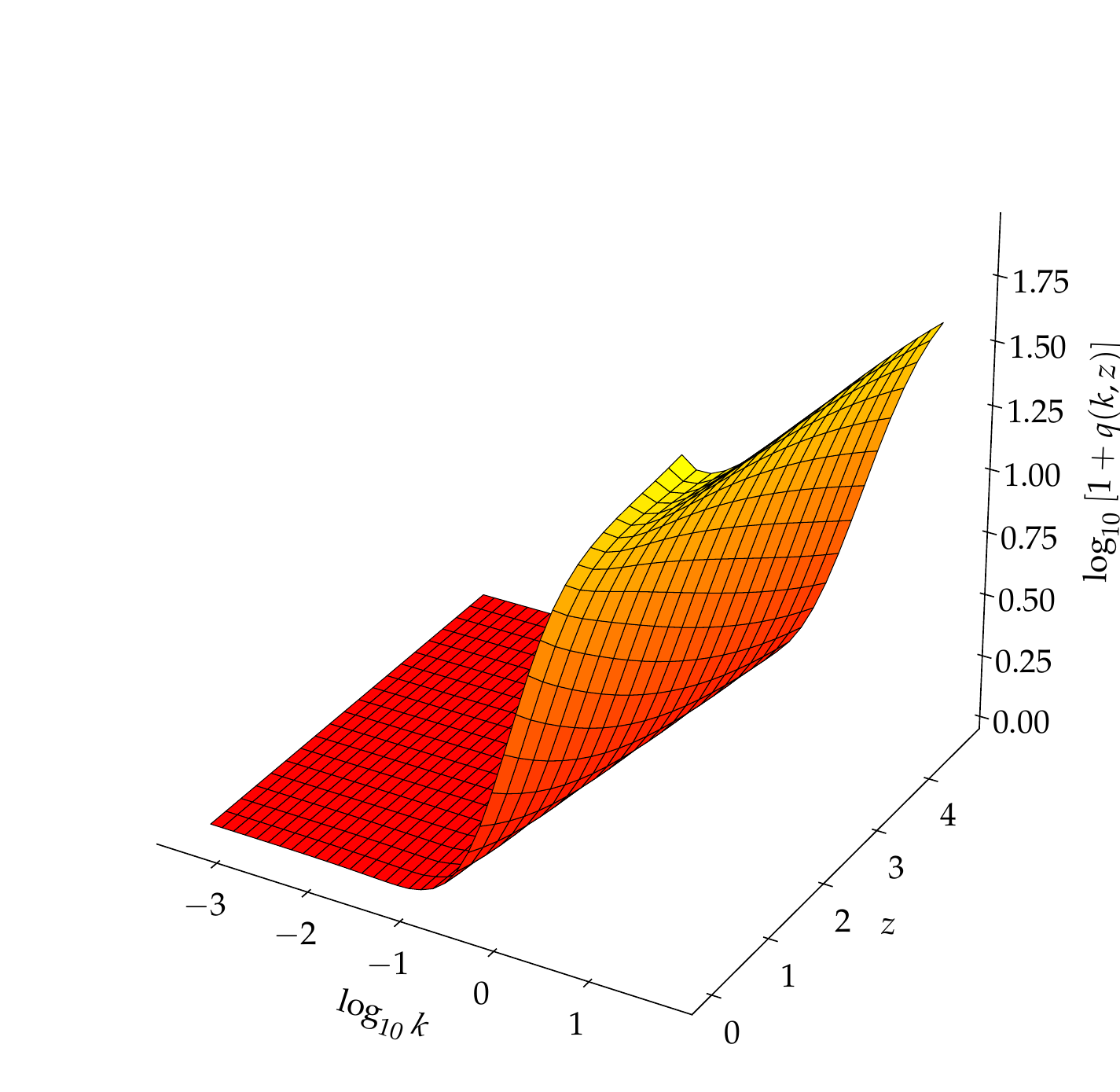}}} 
    \caption{\label{fig:3d_pk_and_q}The left panel shows the 3D matter power spectrum at a fixed input cosmology to CLASS for $k\in[5\times 10^{-4}, 50]$ and $z\in[0.0, 4.66]$. The grid shows the region where we choose to model the function, that is, 40 wavenumbers, equally spaced in logarithm scale and 20 redshifts, equally spaced in linear scale.}
\end{figure*}

The ML techniques discussed previously will slowly pave their way in various weak lensing (WL) analysis. Indeed, in the analysis of the cosmic microwave background (CMB), \cite{2007ApJ...654....2F} designed the Parameters for the Impatient Cosmologist (PICO) algorithm for interpolating CMB power spectra at test points in the parameter space. In the same spirit, \cite{2007MNRAS.376L..11A} built a neural network algorithm, which they refer to as CosmoNet, for interpolating CMB power spectra. Neural networks have been used in other applications as well, for example, in simulations. \cite{2012MNRAS.424.1409A, 2014MNRAS.439.2102A} used neural networks for interpolating non-linear matter power spectrum based on 6 cosmological parameters while \cite{2018MNRAS.475.1213S} used neural networks for emulating the 21cm power spectrum in the context of epoch of reionisation. In the context of weak lensing analysis, \cite{2020MNRAS.491.2655M} used neural networks for accelerating cosmological parameter inference by combining cosmic shear, galaxy clustering, and tangential shear. While we were finishing this work, the work of \cite{2021arXiv210414568A} and \cite{2021arXiv210501081H}, both related to emulating the matter power spectrum, appeared on arXiv. In particular, \cite{2021arXiv210501081H} used GPs to build an emulator for the matter power spectrum at fixed redshifts using N-body simulations while \cite{2021arXiv210414568A} used neural networks and a combination of LH points (an 8D input parameter space with 156\,000 training points), which they refer to as the standard (10$\sigma$ around Planck best-fitting parameters) and extended (roughly twice as large as the standard one) cosmological space to emulate the linear matter power spectrum as well as other cross-spectra of linear fields. \citet{2021arXiv210603846S} also used neural networks to emulate the 3D non-linear matter power spectrum, with at least $10^{5}$ training points depending on their applications and the redshift is also treated as an input to the neural network.  It is possible to train neural networks to return the derivatives with respect to both the parameters (weights and biases) and the inputs, but this is not done in current implementations.  With GPs, the derivatives are trivially obtained analytically, without any training on potentially noisy numerical derivatives. \citet{2019JCAP...09..028A} also used neural networks for approximating CMB power spectra.

On the other hand, Gaussian Processes have been used in the Coyote Universe collaboration \citep{2007PhRvD..76h3503H, 2009ApJ...705..156H, 2010ApJ...715..104H, 2014ApJ...780..111H, 2010ApJ...713.1322L} for emulating the matter power spectrum for large-scale simulations. Recently, \cite{2018PhRvD..98f3511L} used Gaussian Processes in the context of likelihood-free inference, where the data (training points) is augmented in an iterative fashion via Bayesian Optimisation, hence the procedure being referred to as Bayesian Optimisation for Likelihood-Free Inference, BOLFI \citep{JMLR:v17:15-017}. Each emulating scheme has its own pros and cons (we defer to \S\ref{sec:software} for a short discussion on the advantages and possible limitations of Gaussian Processes).

Different emulating schemes have been designed for the matter power spectrum and most of them are based on combining Singular Value Decomposition (SVD) and Gaussian Processes. The emulator from \cite{2007PhRvD..76h3503H} is among the first in the context of large simulations. Emulating $P_{\delta}(k,z)$ is not a trivial task because it is strictly a function of 3 inputs, $k$, the wavenumber, $z$, the redshift and $\bs{\theta}$, the cosmological parameters. Neural networks seem to be the obvious choice because they can deal with multiple outputs but they generally require a large number of training points. 

Our contributions in this work are three fold. First, it addresses the point that we do not always need to assume a zero mean Gaussian Process model for performing emulation, in other words, one can also include some additional basis functions prior to defining the kernel matrix. This can be useful if we already have an approximate model of our function. Moreover, if we know how a particular function behaves, one can adopt a stringent prior on the regression coefficients for the parametric model, hence allowing us to encode our degree of belief about that specific parametric model. Second, since we are using a Radial Basis Function (RBF) kernel and the fact that it is infinitely differentiable enables us to estimate the first and second derivatives of the 3D matter power spectrum. The derived expressions for the derivatives also indicate that there is only element-wise matrix multiplication and no matrix inverse to compute. This makes the gradient calculations very fast. Finally, with the approach that we adopt, we show that the emulator can output various key power spectra, namely, the linear matter power spectrum at a reference redshift $z_{0}$ and the non-linear 3D matter power spectrum with/without an analytic baryon feedback model. Moreover, using the emulated 3D power spectrum and the tomographic redshift distributions, we also show that the weak lensing power spectrum and the intrinsic alignment (II and GI) can be generated in a very fast way using existing numerical techniques. 

\begin{figure*}
    \centering
    \subfloat[The growth factor, $D(z)$]{{\includegraphics[width=0.40\textwidth]{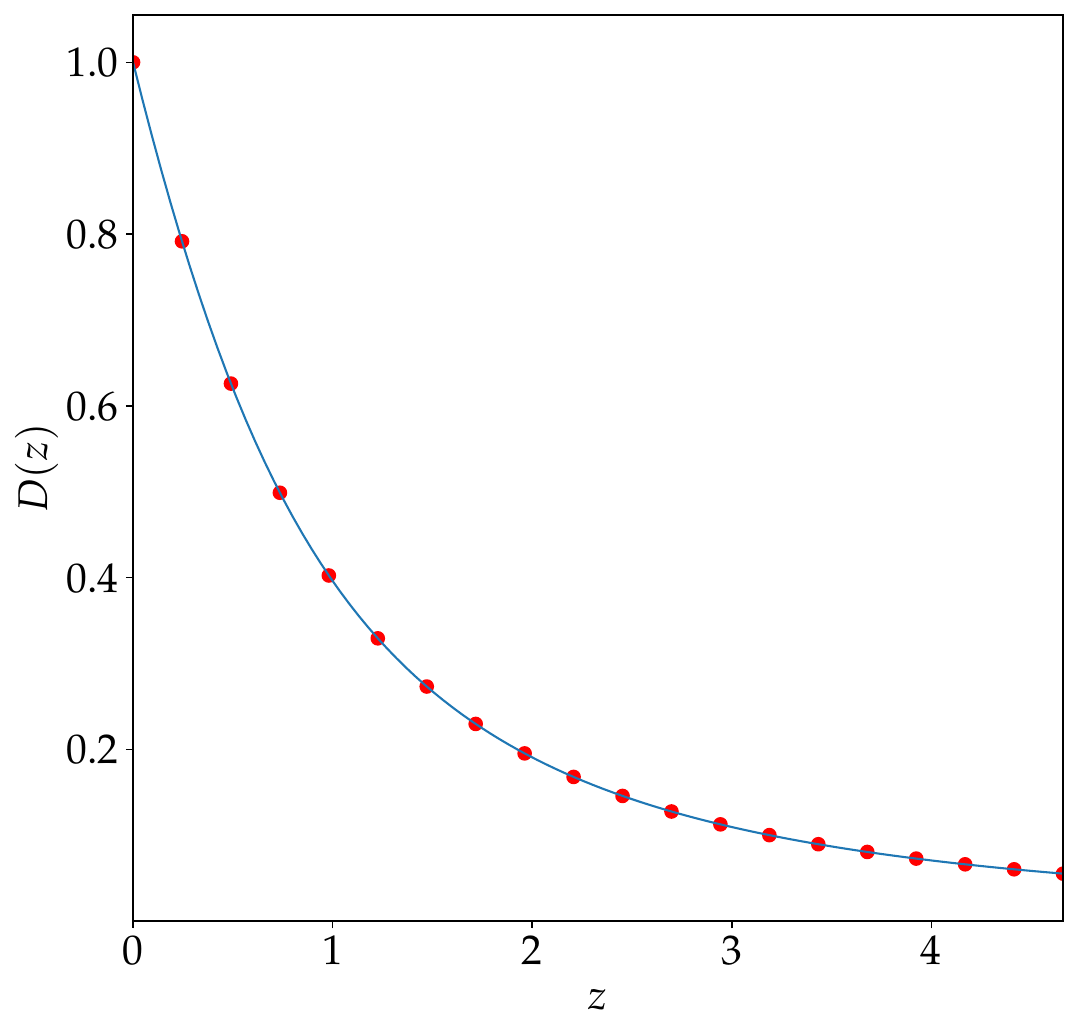}}}
    \qquad
    \subfloat[The linear and non-linear matter power spectrum]{{\includegraphics[width=0.40\textwidth]{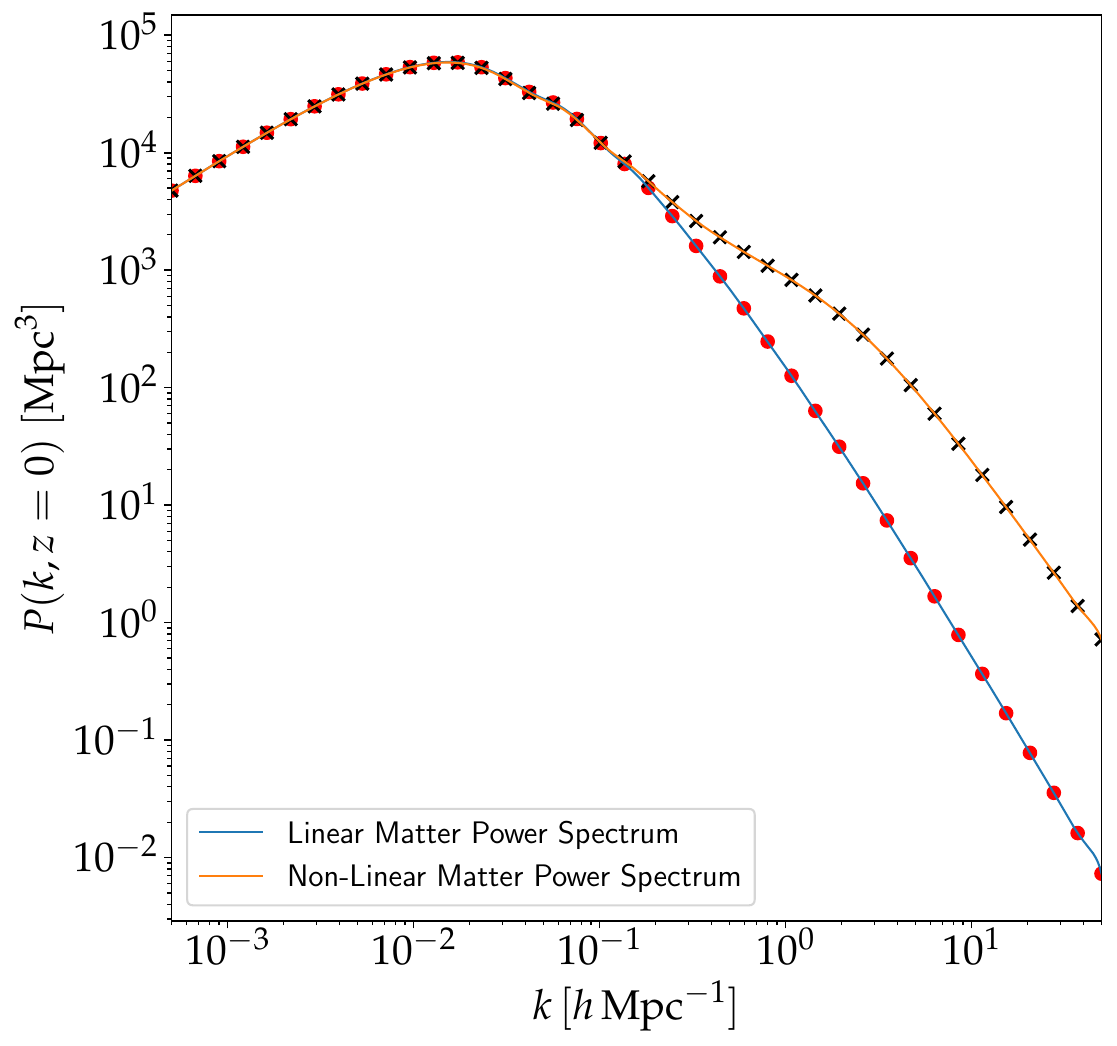}}} 
    \caption{\label{fig:gf_pk_lin}The left panel shows the growth factor, as function of redshift. In this case, to generate the training set, the growth factor is calculated at 20 redshifts, equally spaced in linear scale (shown by the red scatter points) and the linear matter power spectrum, $P_{\tm{lin}}(k,z_{0})$ is calculated at 40 different wavenumbers, $k$, equally spaced in logarithm space (red scatter points). We also show the non-linear matter power spectrum in (b). These functions are evaluated at different cosmological parameters to build a training set.}
\end{figure*}

In \cite{2020MNRAS.497.2213M}, we found that using the mean of the GP and ignoring the error always results in better posterior densities. This is a known feature when GPs emulate a deterministic function \citep{doi:10.1198/TECH.2009.08019}. As a result, in this work, we work only with the mean of the GP in all experiments. We use 1000 training points and once the emulator is trained and stored, it takes about 0.1 seconds to generate the non-linear 3D matter power spectrum, compared to CLASS which takes about 30 seconds to generate an accurate and smooth power spectrum. Hence, the method presented in this paper also opens a new avenue towards building emulators for large-scale simulations where a single high-resolution forward simulation might take minutes to compute.

The paper is organised as follows: in \S\ref{sec:model}, we describe the 3D power spectrum, which can be decomposed in different components, and the analytic baryon feedback model, which can be used in conjunction with $P_{\delta}(k,z)$. In \S\ref{sec:procedures} and \S\ref{sec:gradient}, we provide a mathematical description for calculating multiple important quantities for the emulator, for example, making predictions at test points, learning the kernel hyperparameters and computing derivatives. In \S\ref{sec:wl}, using a pair of toy $n(z)$ tomographic redshift distributions, we show how the emulator can be used to generate different weak lensing power spectra and in \S\ref{sec:software}, we describe briefly the different functionalities that the code supports and we highlight the main results in \S\ref{sec:results}. Finally, we conclude in \S\ref{sec:conclusions}.

%% file: Content/model.tex
\section{Model}
\label{sec:model}

In this section, we describe the model which we want to emulate. Central to the calculation is the 3D matter power spectrum, $P_{\delta}(k,z;\bs{\theta})$, where $\bs{\theta}$ refers to a vector of cosmological parameters. In what follows, we will drop the $\bs{\theta}$ vector notation for clarity. The matter power spectrum is generally the most expensive part to calculate, especially if one chooses to use large-scale simulation to generate the 3D matter power spectrum. In the simple case, one can just emulate $P_{\delta}(k,z)$ but we consider a different approach, which enables us to include baryon feedback, to calculate the linear matter power spectrum at a reference redshift and to calculate the non-linear 3D matter power spectrum itself. 

Baryon feedback is one of the astrophysical systematics which is included in a weak lensing analysis. This process is not very well understood but is deemed to modify the matter distribution at small scales, hence resulting in the suppression of the matter power spectrum at large multipoles. In general, to model these effects, large hydrodynamical simulations provide a proxy to model baryon feedback. In particular, it is quantified via a bias function, $b^{2}(k,z)$ such that the resulting modified 3D matter power spectrum can be written as

\begin{equation}
P_{\delta}^{\tm{bary}}(k,z)=b^{2}(k,z)P_{\delta}(k,z),
\end{equation}

\noindent where $P_{\delta}^{\tm{bary}}(k,z)$ and $P_{\delta}(k,z)$ are the 3D matter power spectra, including and excluding baryon feedback respectively. The bias function is modelled by the fitting formula

\begin{equation}
b^{2}(k,z)=1-A_{\tm{bary}}\left[A_{z}e^{(B_{z}x-C_{z})^{3}}-D_{z}xe^{E_{z}x}\right]\;,
\end{equation}

\noindent where $A_{\tm{bary}}$ is a flexible nuisance parameter and we allow it to vary over the range $A_{\tm{bary}}\in[0.0,2.0]$. The quantity $x=\tm{log}_{10}(k\;[\tm{Mpc}^{-1}])$, and $A_{z}$, $B_{z}$, $C_{z}$, $D_{z}$ and $E_{z}$ depend on the redshift and other constants. See \cite{2015MNRAS.450.1212H} for details and functional forms. Note that setting $A_{\tm{bary}}=0$ implies no baryon feedback. Moreover, since we have a functional form for the baryon feedback model, which is not expensive to compute, we will apply it as a bolt-on function on top of the emulated non-linear 3D matter power spectrum.

Next, we consider the non-linear 3D matter power spectrum without baryon feedback. It can be decomposed into three components as follows:

\begin{equation}
    P_{\delta}(k,z)=D(z)[1+q(k,z)]P_{\tm{lin}}(k,z_{0})
\end{equation}

\noindent where $D(z)$ is the linear growth factor (assumed scale-independent), and $P_{\tm{lin}}(k,z_{0})$ is a scale-independent reference linear matter power spectrum at fixed redshift $z_{0}$. The quantity $q(k,z)$ not only encapsulates the non-linear contributions, but also any scale-dependence in the linear growth factor, for instance due to massive neutrinos or modified gravity. See Figures \ref{fig:3d_pk_and_q} and \ref{fig:gf_pk_lin} for an illustration of the decomposition of the 3D matter power spectrum at fixed cosmological parameters. Emulating the three different components separately has the advantage of calculating the linear matter power spectrum at the reference redshift for any given input cosmology.

Following current weak lensing analysis, we define some bounds on the redshifts, $z$ and wavenumbers, $k$. For example, the maximum redshift in the tomographic weak lensing analysis performed by \cite{2017MNRAS.471.4412K} is $\sim 5$ and the maximum wavenumber is set to 50. With these numbers in mind, we choose $z\in[0.0, 5]$ and $k\in[5\times 10^{-4}, 50]$. We will elaborate more on these settings in the sections which follow. On the other hand, for the cosmological parameters, we assume the following range to generate the training set: 

\begin{table}[H]
\footnotesize
\caption{\label{tab:prior_range}Default parameter prior range inputs to the emulator}
\renewcommand\arraystretch{1.5}
\noindent \begin{centering}
\begin{tabularx}{0.45\textwidth} { 
  | >{\hsize=0.3\textwidth}X 
  | >{\hsize=0.1\textwidth}X | }
\hline
\textbf{Description} & \textbf{Range}\tabularnewline
\hline 
CDM density, $\Omega_{\tm{cdm}}h^{2}$ & $[0.06,\,0.40]$\tabularnewline
Baryon density, $\Omega_{\tm{b}}h^{2}$ & $[0.019,\,0.026]$\tabularnewline
Scalar spectrum amplitude, $\tm{ln}(10^{10}A_{s})$ & $[1.70,\,5.0]$\tabularnewline
Scalar spectral index, $n_{s}$ & $[0.7,\,1.3]$\tabularnewline
Hubble parameter, $h$ & $[0.64,\,0.82]$\tabularnewline
\hline
\end{tabularx}
\par\end{centering}
\end{table}

Current weak lensing analyses also assume a fixed sum of neutrino mass, $\Sigma m_{\nu}$. Hence, in all experiments, $\Sigma m_{\nu}=0.06\,\tm{eV}$. This quantity can be fixed by the user prior to running all experiments with the pipeline we have developed. However, we can also treat it as a varying parameter before building the emulator. 

%% file: Content/procedures.tex
\section{Procedures}
\label{sec:procedures}
In the existing likelihood code from \cite{2017MNRAS.471.4412K}, the accurate solver, CLASS, is queried at 39 wavenumbers $k$ and 72 redshifts $z$, corresponding to the centres of each tophat in the $n(z)$ distribution and a standard spline interpolation is carried out along the $k$ axis. Following a similar approach, we choose to have a model of the $P_{\delta}(k,z)$ at 40 values of $k$, equally spaced on a logarithmic grid and 20 values of redshift, equally spaced in linear scale from 0 to 4.66 (the maximum redshift in the KiDS-450 analysis) and we can perform a standard 2D interpolation, such as spline interpolation, along $k$ and $z$. See Figures \ref{fig:3d_pk_and_q} and \ref{fig:gf_pk_lin} for an illustration.

In this section, we will walk through the steps to build a model for the 3D matter power spectrum. It is organised as follows: in \S\ref{sec:training_points} we discuss how the input training points are generated and this is crucial for the emulator to work with a reasonable number of training points. In \S\ref{sec:polynomial_regression}, we cover briefly the standard approach of emulating functions via polynomial regression and in \ref{sec:model_residuals}, we elaborate on how we can model the residuals, that is, the discrepancy between the actual function and assumed polynomial function. We denote the response (or target), that is, the function we want to model as $y$. In this particular case, we have three different components, namely the growth factor, $D(z)$, the $q(k,z)$ function and the linear matter power spectrum $P_{\tm{lin}}(k,z_{0})$. We assume we have run the simulator, CLASS, at $N$ design points, $\bs{\theta}$, such that we have a training set, $\{\bs{\theta}, \bs{y}_{i}\}$. Throughout this work, we use the fitting function Halofit \citep{2012ApJ...761..152T} implemented in CLASS to generate the training set. The index $i$ corresponds to the $i^{\tm{th}}$ response. Note that in our application, we model each function independently with the emulating scheme proposed below. 

\subsection{Training Points}
\label{sec:training_points}
An important ingredient in designing a robust emulator lies in generating the input training points. Points which are drawn randomly and uniformly from the pre-defined range (see Table \ref{tab:prior_range}) do not show a space-filling property. As the dimensionality of the problem increases, the emulator may lack training information in its neighbourhood and the prediction can be very poor in these regions. Moreover, one would need a large number of training points to accurately model the power spectrum. For example, a recent work by \citet{2021arXiv210603846S} shows that one would need $\sim 10^{5}$ training points to build an emulator with deep neural networks with uniform random sampling.

\begin{figure}[H]
\noindent \begin{centering}
\includegraphics[width=0.4\textwidth]{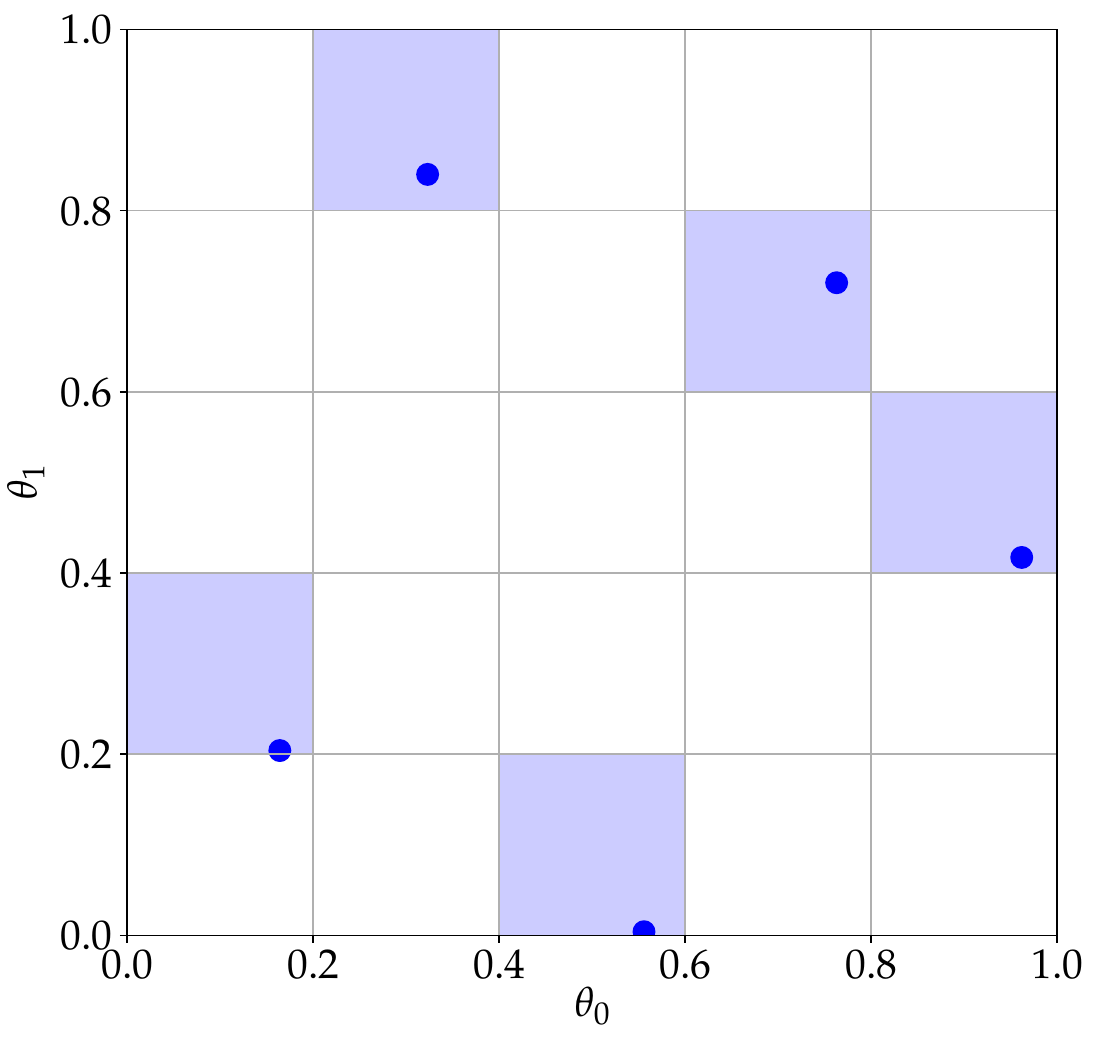}
\par\end{centering}
\caption{\label{fig:lhs_illutrate}An example of a Latin Hypercube (LH) design in two dimensions. Five LH points are drawn randomly using the \texttt{maximin} procedure and each point will occupy a single cell, that is, if a point occupies cell $(i,j)$, then there is not a point occupying cell $(j,i)$. This procedure remains exactly the same when we generate LH samples from a hypercube.}   
\end{figure}

To circumvent these issues, the natural choice is to instead generate Latin Hypercube (LH) samples, which demonstrate a space-filling property as shown in Figure \ref{fig:lhs_illutrate}. The idea behind a LH design is that a point will always occupy a single cell. For example, if we consider the design shown in Figure \ref{fig:lhs_illutrate}, each column and row contains precisely one training point (in 2D). Similarly, in a 3D case, each row, column and layer will have one training point and this extends to higher dimensions. Intuitively, for a 2D design, this is analogous to the problem of positioning $n$ rooks on an $n\times n$ chessboard such that they do not attack each other. This ensures that the LH points generated cover the parameter space as much as possible, hence enabling the emulator to predict the targets at test points. \texttt{emuPK} can also be trained on a different set of training points, for example, one which has been generated using a different LH sampling scheme.

\begin{table}[H]
\footnotesize
\caption{\label{tab:notations}Symbols and notations with corresponding meanings}
\renewcommand\arraystretch{1.5}
\noindent \begin{centering}
\begin{tabularx}{0.45\textwidth} { 
  | >{\hsize=0.06\textwidth}X 
  | >{\hsize=0.34\textwidth}X | }
\hline
\textbf{Symbol} & \textbf{Meaning}\tabularnewline
\hline 
$N$ & Number of training points\tabularnewline
$m$ & Number of basis functions\tabularnewline
$\bs{y}$ & Target of size $N$\tabularnewline
$\bs{\theta}$ & Inputs to the emulator\tabularnewline
$\bs{\beta}$ & Regression coefficients of size $m$\tabularnewline
$\bs{f}$ & Deterministic error component of size $N$ of the model\tabularnewline
$\bs{\Phi}$ & Design matrix of size $N\times m$\tabularnewline
$\sans{K}$ & Kernel matrix of size $N\times N$\tabularnewline
$\sans{C}$ & Prior covariance matrix of $\bs{\beta}$ of size $m\times m$\tabularnewline
$\bs{\mu}$ & Prior mean of $\bs{\beta}$ of size $m$\tabularnewline
$\sans{D}$ & $\sans{D}=\left[\bs{\Phi},\,\bb{I}\right]$ is a new design matrix of size $N\times (m+N)$\tabularnewline
$\bs{\alpha}$ & $\bs{\alpha}=\left[\bs{\beta},\,\bs{f}\right]^{\tm{T}}$ is a vector of size $m+N$\tabularnewline
$\sans{R}$ &  Prior covariance matrix of size $(m+N)\times(m+N)$\tabularnewline
$\bs{\gamma}$ & $\bs{\gamma}=\left[\bs{\mu},\,\bs{0}\right]^{\tm{T}}$ prior mean of size $m+N$\tabularnewline
$\bs{\Sigma}$ & Noise covariance matrix of size $N\times N$\tabularnewline
$d$ & Dimension of the problem\tabularnewline
$\bs{\nu}$ & Kernel hyperparameters \tabularnewline 
\hline
\end{tabularx}
\par\end{centering}
\end{table}

In this application, we use the \texttt{LHS} package, available in \texttt{R} to generate the design points. Whilst many different functions are available to generate the LH samples, we choose the \texttt{maximinLHS} procedure, which maximises the minimum distance between the LH points. If we have a set of design points $(x_{i},y_{i})$ where $x_{i}\neq x_{j}$, $y_{i}\neq y_{j}$ and $i\neq j$, in the case of a maximin LH design, for a certain distance measure, $d$, the separation distance, $\tm{min}_{i\neq j}\;d[(x_{i},y_{i}), (x_{j},y_{j})]$ is maximal \citep{van2007maximin}.

These design points as generated from the \texttt{maximinLHS} procedure lie between 0 and 1 and hence, they are scaled according the range we want to distribute them. For example, if $\theta_{\tm{min}}$ and $\theta_{\tm{max}}$ are the minimum and maximum of a particular parameter, the LH points are scaled as: $\theta=\theta_{\tm{min}}+r(\theta_{\tm{max}}-\theta_{\tm{min}})$, where $r$ is the LH design point. Alternatively, if we want them to follow a specific distribution, for example, a Gaussian distribution, one can just use the inverse cumulative density function to scale the LH point. 

As discussed by \cite{2007ApJ...654....2F}, we also want to ensure for the fact that there is roughly an equal variation in the power spectrum when we take a step in parameter space. This condition can be met by pre-whitening the input parameters prior to building the emulator. This can be achieved as follows: the training points are first centred on 0, that is, $\bs{\theta}'\rightarrow \bs{\theta}-\bar{\bs{\theta}}$. The covariance, $\sans{M}$ of this modified training set is computed, and Cholesky decomposed to $\sans{M}=\sans{L}\sans{L}^\tm{T}$, then $\bs{\theta}'=\sans{L}\tilde{\bs{\theta}}$, where $\tilde{\bs{\theta}}$ has a covariance matrix equal to the identity.

Once we have our training set, our goal is to learn the functional relationship between the function $\bs{y}$ (we have dropped the index $i$ but the same steps apply to the other functions) and the inputs $\bs{\theta}$. In other words, we model the data (simulations), $\bs{y}$, as

\begin{equation}
    \bs{y} = \bs{h}(\bs{\theta}) + \bs{\epsilon}
\end{equation}

\noindent where $\bs{h}$ is the underlying assumed model. The output of CLASS at the training points would often be called ‘data’ in a ML context. Conceptually, this fitting procedure is analogous to many parameter inference tasks in Cosmology, where $\bs{y}$ would be a set of data from observations, for example, a set of band powers and $\bs{h}$ would be a $\Lambda\tm{CDM}$ model.

\subsection{Polynomial Regression}
\label{sec:polynomial_regression}
In our application, $\bs{h}$ might be a deterministic function but the functional (parametric) form might be unknown to us. A straightforward approach is to assume a polynomial approximation to the data, that is, 

\begin{equation}
\label{eq:polynomial_model}
    \bs{y} = \bs{\Phi}\bs{\beta} + \bs{\epsilon},
\end{equation}
\noindent where $\bs{\Phi}$ is a design matrix, whose columns contain the basis functions $[1, \bs{\theta}_{1},\ldots\bs{\theta}_{p}^{n}]$ and $n$ is the order of the polynomial. $\bs{\beta}$ is a vector of regression coefficients (also referred to as weights) and $\bs{\epsilon}$ is the noise vector and $\tm{cov}(\bs{\epsilon})=\bs{\Sigma}$. Using Bayes' theorem, the full posterior distribution of the weights is 

\begin{equation}
p(\bs{\beta}\left|\bs{y}\right.)=\dfrac{p(\bs{y}\left|\bs{\beta}\right.)p(\bs{\beta})}{p(\bs{y})}.    
\end{equation}

\noindent $p(\bs{\beta}\left|\bs{y}\right.)$ is the posterior distribution of $\bs{\beta}$, $p(\bs{y}\left|\bs{\beta}\right.)$ is the likelihood of the data, $p(\bs{\beta})$ is the prior for $\bs{\beta}$ and $p(\bs{y})$ is the marginal likelihood (Bayesian evidence) which does not depend on $\bs{\beta}$. In what follows, the notation $\mc{N}(\bs{x}\left|\right.\bs{\mu},\sans{C})$ denotes a multivariate normal distribution with mean $\bs{\mu}$ and covariance $\sans{C}$.

Assuming a Gaussian likelihood for the data, $\mc{N}(\bs{y}\left|\right.\bs{\Phi}\bs{\beta},\bs{\Sigma})$ and a Gaussian prior for the weights, $\mc{N}(\bs{\beta}\left|\right.\bs{\mu},\sans{C})$, the posterior distribution of $\bs{\beta}$ is another Gaussian distribution, $\mc{N}(\bs{\beta}\left|\right.\bar{\bs{\beta}},\bs{\Lambda})$ with mean and covariance given respectively by

\begin{equation}
\begin{split}
\label{eq:post_poly}
    \bar{\bs{\beta}}&=\bs{\Lambda}(\bs{\Phi}^{\tm{T}}\bs{\Sigma}^{-1}\bs{y}+\sans{C}^{-1}\bs{\mu})\\
    \bs{\Lambda}&=(\sans{C}^{-1}+\bs{\Phi}^{\tm{T}}\bs{\Sigma}^{-1}\bs{\Phi})^{-1}.
\end{split}
\end{equation}

In general, we are also interested in learning the (posterior) predictive distribution at a given test point $\bs{\theta}_{*}$, that is, $p(y_{*}\left|\right.\bs{y},\bs{\theta}_{*})$ and this is another Gaussian distribution,

\begin{equation}
    p(y_{*}\left|\right.\bs{y},\bs{\theta}_{*}) = \mc{N}(y_{*}\left|\right. \bs{\Phi}_{*}\bar{\bs{\beta}},\sigma^{2}_{*}+\bs{\Phi}_{*}\bs{\Lambda}\bs{\Phi}_{*}^{\tm{T}})
\end{equation}

\noindent where $y_{*}$ is the predicted function and $\bs{\Phi}_{*}$ is the set of basis functions evaluated at the test point. For noise-free regression, the noise variance, $\sigma^{2}_{*}\approx 0$ and the predictive uncertainty is dominated by the term $\bs{\Phi}_{*}\bs{\Lambda}\bs{\Phi}_{*}^{\tm{T}}$. Moreover, in practice, the noise term at the test point is barely known and is hence approximated by $\bs{\Phi}_{*}\bs{\Lambda}\bs{\Phi}_{*}^{\tm{T}}$. 

On the other hand, we are also interested in understanding the model, that is, the number of basis functions we would need to fit the data. An important quantity is the marginal likelihood which penalises model complexity \citep{1996ApJ...471...24J, doi:10.1080/00107510802066753}. In this case, this quantity can be analytically derived and is given by 

\begin{equation}
    p(\bs{y})=\mc{N}(\bs{y}\left|\right.\bs{\Phi}\bs{\mu},\,\bs{\Sigma}+\bs{\Phi}\sans{C}\bs{\Phi}^{\tm{T}}).
\end{equation}

\noindent Note that this quantity is independent of $\bs{\beta}$ and is an integral of the numerator with respect to \textit{all} the variables (in our case $\bs{\beta}$), that is, 

\begin{equation}
p(\bs{y})=\int p(\bs{y}\left|\bs{\beta}\right.)p(\bs{\beta})\; \tm{d}\bs{\beta}.
\end{equation}

\noindent To this end, one can compute the Bayesian evidence for a series of (polynomial) models and choose the model which yields the maximum Bayesian evidence \citep{2006PhRvD..74b3503K}. 

\subsection{Modelling the residuals}
\label{sec:model_residuals}
The above formalism works well in various cases but (1) polynomial model fitting is generally a \textit{global} fitting approach, (2) there exists a large number of choice for the number of basis functions, and (3) the functional relationship between the data and the model might be a very complicated function. In this section, we therefore propose a Bayesian technique which models the residuals, that is, the difference between our proposed polynomial approximation and the underlying model. We will re-write equation (\ref{eq:polynomial_model}) as 

\begin{equation}
\label{eq:parametric_gp}
    \bs{y} = \bs{\Phi}\bs{\beta} + \bs{f} + \bs{\epsilon}, 
\end{equation}

\noindent where $\bs{f} = \bs{h} - \bs{\Phi}\bs{\beta}$ is the deterministic error component of the model \citep{10.1093/biomet/62.1.79}. Under the assumption that we have modelled $\bs{y}$ as much as we can with the polynomial model, it is fair to make an a priori assumption for the distribution of $f$. In function space, points which are close to each other will depict similar values for $f$ and as we move further away from a given design point, it is expected that the degree of similarity will decrease. In other words, the correlation between $f(\bs{\theta}_{i})$ and $f(\bs{\theta}_{j})$ decreases monotonically as the distance between $\bs{\theta}_{i}$ and $\bs{\theta}_{j}$ increases. This prior knowledge can be encapsulated by using a covariance (kernel) function such as the Gaussian function, that is, 

\begin{equation}
    \tm{cov}(f_{i},f_{j}) = \lambda^{2}\tm{exp}\left[-\dfrac{1}{2}(\bs{\theta}_{i}-\bs{\theta}_{j})^{\tm{T}}\bs{\Omega}^{-1}(\bs{\theta}_{i}-\bs{\theta}_{j})\right], 
\end{equation}

\noindent where $\bs{\Omega}=\tm{diag}(\omega^{2}_{1}\ldots\omega^{2}_{d})$ and $\omega^{2}_{i}$ is the characteristic lengthscale for each dimension. $\bs{\nu}=\{\lambda,\omega_{1},\ldots\omega_{d}\}$ is the set of hyperparameters for this kernel. In the same spirit, the full prior distribution for $\bs{f}$ is a multivariate normal distribution, that is, 

\begin{equation}
    p(\bs{f})=\mc{N}(\bs{f}\left|\right.\bs{0},\sans{K})
\end{equation}

\noindent where the kernel matrix has elements $k_{ij}\equiv\tm{cov}(f_{i},f_{j})$. At this point, we will assume that the hyperparameters are fixed but we will later consider learning them via optimisation. 

\subsubsection{Inference}
Now that we have a model for the data (training set), we seek the full posterior distribution of the variables $\bs{\beta}$ and $\bs{f}$. We assume a Gaussian prior for $\bs{\beta}$, that is, $p(\bs{\beta})=\mc{N}(\bs{\beta}\left|\right.\bs{\mu},\sans{C})$. Using Bayes' theorem, the posterior distribution of $\bs{\beta}$ and $\bs{f}$ is

\begin{equation}
    p(\bs{\beta},\bs{f}\left|\right.\bs{y})=\dfrac{p(\bs{y}\left|\right.\bs{\beta},\bs{f})p(\bs{\beta},\,\bs{f})}{p(\bs{y})}
\end{equation}

\noindent To simplify the derivation, we will rewrite equation (\ref{eq:parametric_gp}) as

\begin{equation}
    \bs{y} = \sans{D}\bs{\alpha} + \bs{\epsilon},
\end{equation}

\noindent where $\sans{D}=[\bs{\Phi},\bb{I}]$ is an augmented, new design matrix, consisting of the existing design matrix $\bs{\Phi}\in\bb{R}^{N\times m}$ and the identity matrix, $\bb{I}$ of size $N\times N$. $\bs{\alpha} = [\bs{\beta},\bs{f}]^{\tm{T}}$ is now a vector of length $N+m$, consisting of both $\bs{\beta}$ and $\bs{f}$. The sampling distribution of $\bs{y}$ is a Gaussian distribution, $\mc{N}(\bs{y}\left|\right.\sans{D}\bs{\alpha}, \bs{\Sigma})$. We can rewrite the full prior distribution of both set of parameters, $\bs{\beta}$ and $\bs{f}$ as $\mc{N}(\bs{\alpha}\left|\right.\bs{\gamma},\sans{R})$, where

\begin{equation*}
\bs{\gamma}=\left[\begin{array}{c}
\bs{\mu}\\
\bs{0}
\end{array}\right]\;\;\tm{and}\;\;\sans{R}=\left[\begin{array}{cc}
\sans{C} & \sans{0}\\
\sans{0} & \sans{K}
\end{array}\right]
\end{equation*}

Using a similar approach as in the previous section, the posterior of $\bs{\alpha}$ is another Gaussian distribution, that is, 
\begin{equation}
    p(\bs{\alpha}\left|\right.\bs{y})=\mc{N}(\bs{\alpha}\left|\right.\sans{A}^{-1}\bs{b}, \sans{A}^{-1}),
\end{equation}

\noindent where $\sans{A}=\sans{D}^{\tm{T}}\bs{\Sigma}^{-1}\sans{D} + \sans{R}^{-1}$ and $\bs{b}=\sans{D}^{\tm{T}}\bs{\Sigma}^{-1}\bs{y}+\sans{R}^{-1}\bs{\gamma}$. The covariance of $\bs{\beta}$ and $\bs{f}$ are given respectively by:

\begin{equation}
\label{eq:gp_cov}
\sans{V}_{\bs{\beta}} = \left[\bs{\Phi}^{\tm{T}}\left(\sans{K}+\bs{\Sigma}\right)^{-1}\bs{\Phi}+\sans{C}^{-1}\right]^{-1}
\end{equation}

\noindent and 

\begin{equation}
\label{eq:cov_f}
\sans{V}_{\bs{f}} = \left[\sans{K}^{-1} + (\bs{\Sigma}+\bs{\Phi}\sans{C}\bs{\Phi}^{\tm{T}})^{-1}\right]^{-1}
\end{equation}

\noindent Moreover, the posterior mean for $\bs{\beta}$ and $\bs{f}$ can be derived and are given respectively by

\begin{equation}
\label{eq:gp_post_beta}
    \hat{\bs{\beta}} = \sans{V}_{\bs{\beta}}\left[\bs{\Phi}^{\tm{T}}\left(\sans{K}+\bs{\Sigma}\right)^{-1}\bs{y} + \sans{C}^{-1}\bs{\mu}\right]
\end{equation}

\noindent and 

\begin{equation}
\label{eq:gp_post_f}
    \hat{\bs{f}} = \sans{V}_{f}\bs{\Sigma}^{-1}\left[\bs{y} - \bs{\Phi}\bar{\bs{\beta}}\right]
\end{equation}

\noindent Recall that $\bar{\bs{\beta}}$ is the expression for the posterior distribution of $\bs{\beta}$ when we use the polynomial model only. There are also some useful remarks and sanity checks which we can make from equations \ref{eq:gp_cov}, \ref{eq:gp_post_beta} and \ref{eq:gp_post_f}. In equation (\ref{eq:gp_cov}), for the covariance of $\bs{\beta}$, if we had ignored the other variables $\bs{f}$, in other words, in the absence of the kernel matrix, $\sans{K}$, we recover the posterior covariance for $\bs{\beta}$ when we use a polynomial model only. A similar argument applies for equation (\ref{eq:gp_post_beta}) in which case we also recover the posterior distribution of $\bs{\beta}$ in the polynomial model. Equation (\ref{eq:gp_post_f}) has a nice interpretation. The posterior mean of $\bs{f}$ is a linear combination of the residuals, $\bs{y}-\bs{\Phi}\bar{\bs{\beta}}$. 

\subsubsection{Prediction}

Now that we have the full posterior distribution of the variables, another key ingredient is learning the predictive distribution at a given test point, $\bs{\theta}_{*}$. The joint distribution of the data and the function at the test point can be written as 

\begin{equation}
\left[\begin{array}{c}
\bs{y}\\
y_{*}
\end{array}\right]\sim\mc{N}\left(\left[\begin{array}{c}
\bs{\Phi}\bs{\beta}\\
\bs{\Phi}_{*}\bs{\beta}
\end{array}\right],\,\left[\begin{array}{cc}
\sans{K}+\bs{\Sigma} & \bs{k}_{*}\\
\bs{k}_{*}^{\tm{T}} & k_{**}+\sigma_{*}^{2}
\end{array}\right]\right)
\end{equation}

\noindent where $\bs{k}_{*}$ is a vector, whose elements are given by calculating the kernel function between each training point and the test point, $\bs{\theta}_{*}$. Similarly, $k_{**}$ is just the kernel function evaluated at the test point only. The conditional distribution of $y_{*}$ is a Gaussian distribution

\begin{equation}
    p(y_{*}\left|\right.\bs{y}, \bs{\theta}_{*})=\mc{N}(y_{*}\left|\right.\bar{y}_{*},\tm{var}(y_{*}))
\end{equation}

\noindent where $\bar{y}_{*}$ and $\tm{var}(y_{*})$ are the mean and variance given respectively by

\begin{equation}
\label{eq:prediction}
    \begin{split}
        \bar{y}_{*} &= \sans{X}_{*}\hat{\bs{\beta}} + f_{*}\\
        \tm{var}(y_{*}) &= \sans{X}_{*}\sans{V}_{\beta}\sans{X}_{*}^{\tm{T}} + k_{**} + \sigma_{*}^{2} -\bs{k}_{*}^{\tm{T}}\sans{K}_{y}^{-1} \bs{k}_{*}
    \end{split}
\end{equation}

\noindent and we have defined $\sans{K}_{y}=\sans{K}+\bs{\Sigma}$, $\sans{X}_{*}=\bs{\Phi}_{*}-\bs{k}_{*}^{\tm{T}}\sans{K}_{y}^{-1}\bs{\Phi}$ and $f_{*} = \bs{k}_{*}^{\tm{T}}\sans{K}_{y}^{-1}\bs{y}$. This is another interesting result because if we did not have the parametric polynomial model, then the prediction corresponds to that of a zero mean Gaussian Process (GP) \citep{2006gpml.book.....R}. In our application, once we predict the three components, $D(z)$, $q(k,z)$ and $P_{\tm{lin}}(k,z_{0})$ at a test point $\bs{\theta}_{*}$, the 3D power spectrum easily be calculated using

\begin{equation}
P_{\delta}(k_{*},z_{*};\,\bs{\theta}_{*})=D(z_{*};\,\bs{\theta}_{*})[1+q(k_{*},z_{*};\,\bs{\theta}_{*})]P_{\tm{lin}}(k_{*},z_{0};\,\bs{\theta}_{*})
\end{equation}

Until now, we have assumed a fixed set of kernel hyperparameters. In the next section, we will explain how we can learn them via optimisation. 

\subsubsection{Kernel Hyperparameters}
An important quantity in learning the kernel hyperparameters is the marginal likelihood (Bayesian evidence), which is obtained by marginalising over all the variables $\bs{\alpha}$ and is given by 

\begin{equation}
    p(\bs{y}) = \int p(\bs{y}\left|\right.\bs{\alpha})p(\bs{\alpha})\,\tm{d}\bs{\alpha}.
\end{equation}

\noindent Fortunately, the above integration is a convolution of two multivariate normal distributions, $\mc{N}(\bs{y}\left|\sans{D}\bs{\alpha},\,\bs{\Sigma}\right.)$ and $\mc{N}(\bs{\alpha}\left|\bs{\gamma},\,\sans{R}\right.)$ and hence can be calculated analytically, that is, 

\begin{equation}
    p(\bs{y}) = \mc{N}(\bs{y}\left|\right.\bs{\Phi}\bs{\mu},\,\sans{K}_{y}+\bs{\Phi}\sans{C}\bs{\Phi}^{\tm{T}})
\end{equation}

\noindent and the log-marginal likelihood is 

\begin{equation}
\label{eq:marginal_likelihood}
\begin{split}
\tm{log}\,p(\bs{y}) &= -\dfrac{1}{2}(\bs{y}-\bs{\Phi}\bs{\mu})^{\tm{T}}(\sans{K}_{y}+\bs{\Phi}\sans{C}\bs{\Phi}^{\tm{T}})^{-1}(\bs{y}-\bs{\Phi}\bs{\mu})\\
& -\dfrac{1}{2}\tm{log}\left|\sans{K}_{y}+\bs{\Phi}\sans{C}\bs{\Phi}^{\tm{T}}\right| + \tm{constant}.
\end{split}
\end{equation}

The first term in equation (\ref{eq:marginal_likelihood}) encourages the fit to the data while the second term (the determinant term) controls the model complexity. Recall that the kernel matrix, $\sans{K}$ is a function of the hyperparameters $\bs{\nu}=\{\lambda,\omega_{1},\ldots\omega_{d}\}$. We want to maximise the marginal likelihood with respect to the kernel hyperparameters and this step is equivalent to minimising the cost, that is, the negative log-marginal likelihood. In other words, 

\begin{equation}
    \bs{\nu}_{\tm{opt}} = \underset{\bs{\nu}}{\tm{arg min}}\,J(\bs{\nu)}
\end{equation}

\noindent where we have defined $J(\bs{\nu})\equiv -2\tm{log}\,p(\bs{y})$. An important ingredient for the optimisation to perform well is the gradient of the cost with respect to the kernel hyperparameters, which is given by 

\begin{equation}
    \dfrac{\partial J(\bs{\nu})}{\partial \bs{\nu}_{i}} = \tm{tr}\left[\left((\sans{K}_{y}+\bs{\Phi}\sans{C}\bs{\Phi}^{\tm{T}})^{-1}-\bs{\eta}\bs{\eta}^{\tm{T}}\right)\dfrac{\partial\sans{K}}{\partial\bs{\nu}_{i}}\right],
\end{equation}

\noindent where $\bs{\eta}=(\sans{K}_{y}+\bs{\Phi}\sans{C}\bs{\Phi}^{\tm{T}})^{-1}\bs{y}$. There are a few computational aspects which we should consider when implementing this method. In particular, for a single predictive variance calculation (see equation (\ref{eq:prediction})) an $\mc{O}(N^{2})$ operation is required whereas training (that is, learning the kernel hyperparameters) requires an $\mc{O}(N^{3})$ operation. On the other hand, the mean is quick to compute since it involves $\mc{O}(N)$ operation. 

%% file: Content/gradients.tex
\section{Gradients}
\label{sec:gradient}

\begin{figure*}
\noindent \begin{centering}
\includegraphics[width=0.9\textwidth]{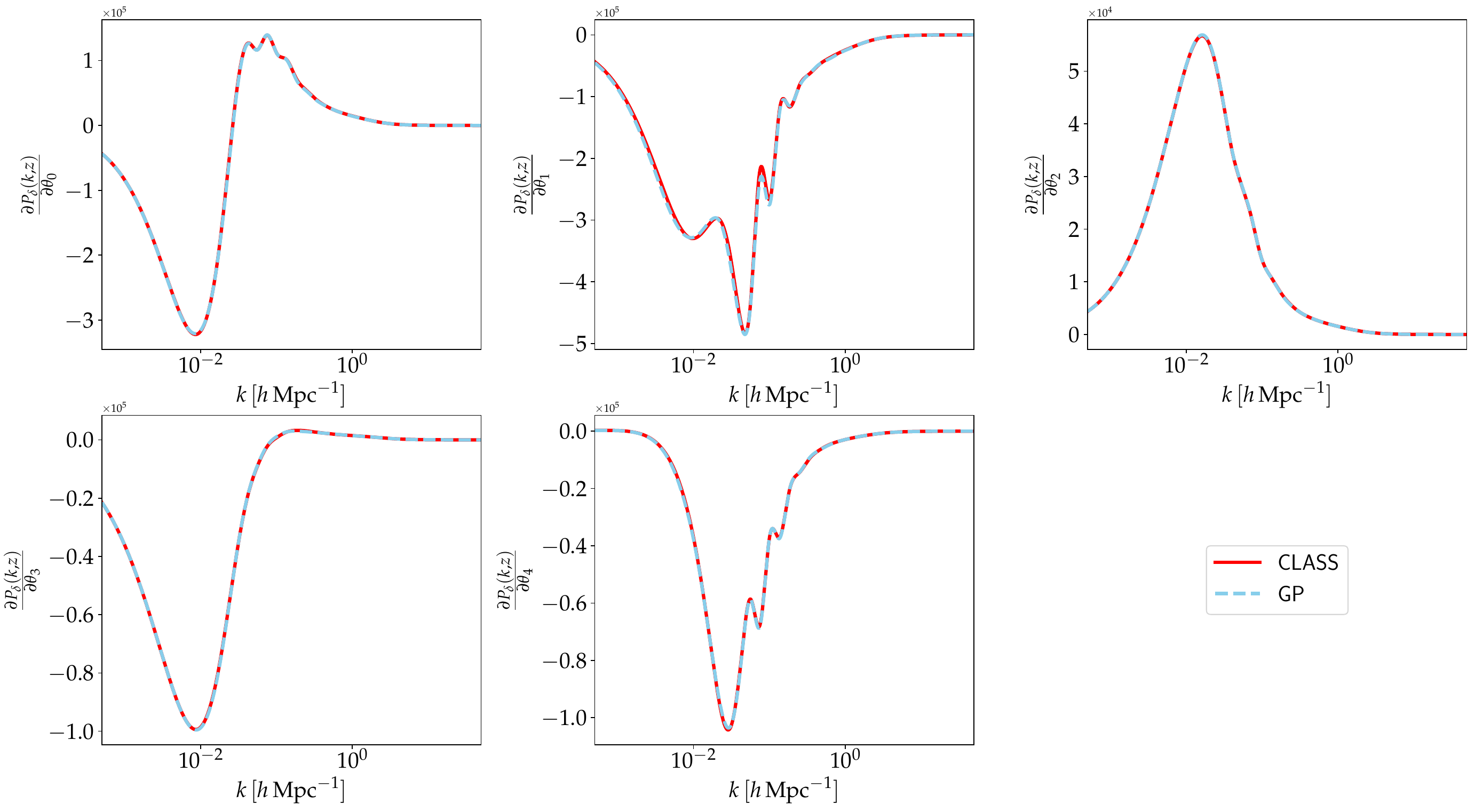}
\par\end{centering}
\caption{\label{fig:gradients}Gradients with respect to the input cosmologies. $\bs{\theta}$ corresponds to the following cosmological parameters: $\bs{\theta}=(\Omega_{\tm{cdm}}h^{2},\,\Omega_{\tm{b}}h^{2},\,\tm{ln}(10^{10}A_{s}),\,n_{s},\,h)$. Note that since we are emulating the 3D power spectrum, the gradient is also a 3D quantity. In this figure, we are showing the predicted function with the GP model in broken blue and the accurate gradient function calculated with CLASS in solid red, at a fixed redshift.}   
\end{figure*}

An important by-product from the trained model is the gradient of the emulated function with respect to the input parameters. This can be of paramount importance if we are using a sophisticated Monte Carlo sampling scheme such as Hamiltonian Monte Carlo (HMC) to infer cosmological parameters in a Bayesian analysis. The gradients of the log-likelihood with respect to the cosmological parameters are important in such a sampling scheme. Hence, with some linear algebra and using the gradient of the power spectra, generated with the emulator, the desired gradients can be derived. The analytical gradient of the mean function with respect to the inputs, at a fixed redshift and wavenumber is

\begin{equation}
\label{eq:gp_grad}
\dfrac{\partial\bar{y}_{*}}{\partial\bs{\theta}_{*}} = \dfrac{\partial\bs{\Phi}_{*}}{\partial\bs{\theta}_{*}}\hat{\bs{\beta}} + \left[\bs{k}_{*}\odot\sans{Z}_{*}\bs{\Omega}^{-1}\right]^{\tm{T}}\sans{K}_{y}^{-1}(\bs{y} - \bs{\Phi}\hat{\bs{\beta}})
\end{equation}

\noindent where $\odot$ refers to element-wise multiplication (Hadamard product). $\sans{Z}_{*}\in\bb{R}^{N\times d}$ corresponds to the pairwise difference between the test point, $\bs{\theta}_{*}$ and the training points, that is, $\sans{Z}_{*}=\left[\bs{\theta}_{1}-\bs{\theta}_{*},\,\bs{\theta}_{2}-\bs{\theta}_{*}\ldots\bs{\theta}_{N}-\bs{\theta}_{*}\right]^{\tm{T}}$. Importantly, as seen from equation \ref{eq:gp_grad}, the gradient is the sum of the gradients corresponding to the parametric part and the residual, which is modelled by a kernel. Moreover, higher order derivatives can also be calculated analytically. For example, the second order auto- and cross- derivatives are

\begin{equation}
    \dfrac{\partial^{2}\bar{y}_{*}}{\partial\bs{\theta}^{2}_{*}}=\dfrac{\partial^{2}\bs{\Phi}_{*}}{\partial\bs{\theta}^{2}_{*}}\bs{\hat{\beta}}+\left[\Omega^{-1}\dfrac{\partial \bs{k}_{*}}{\partial\bs{\theta}_{*}}\sans{Z}_{*}-\Omega^{-1}\odot\bs{k}_{*}\right]\sans{K}_{y}^{-1}(\bs{y} - \bs{\Phi}\hat{\bs{\beta}}).
\end{equation}

\noindent As a result of this procedure, one can analytically calculate the first and second derivatives of an emulated function using kernel methods. While the first derivatives are particularly useful in HMC sampling method, the second derivatives are more relevant in the calculation of, for example, the Fisher information matrix.

Once the gradients with respect to each component of the non-linear 3D matter power spectrum are derived, the first and second derivatives with respect to the non-linear matter spectrum can be derived via chain rule and are given by:

\begin{equation}
\dfrac{\partial P_{\delta}}{\partial\boldsymbol{\theta}}=\dfrac{\partial D}{\partial\boldsymbol{\theta}}(1+q)P_{\textrm{lin}}+D\dfrac{\partial q}{\partial\boldsymbol{\theta}}P_{\textrm{lin}}+D(1+q)\dfrac{P_{\textrm{lin}}}{\partial\boldsymbol{\theta}}  
\end{equation}

\noindent and 

\begin{equation}
\begin{split}
\dfrac{\partial^{2} P_{\delta}}{\partial\boldsymbol{\theta}^{2}} &= \dfrac{\partial^{2} D}{\partial \boldsymbol{\theta}^{2}}(1+q)P_{\textrm{lin}}+D\dfrac{\partial^{2} q}{\partial\boldsymbol{\theta}^{2}}P_{\textrm{lin}}+D(1+q)\dfrac{\partial^{2}P_{\textrm{lin}}}{\partial\boldsymbol{\theta}^{2}}\\
&+2\dfrac{\partial D}{\partial \boldsymbol{\theta}}\dfrac{\partial q}{\partial \boldsymbol{\theta}}P_{\textrm{lin}}+2\dfrac{\partial D}{\partial\boldsymbol{\theta}}(1+q)\dfrac{\partial P_{\textrm{lin}}}{\partial\boldsymbol{\theta}}+2D\dfrac{\partial q}{\partial\boldsymbol{\theta}}\dfrac{\partial P_{\textrm{lin}}}{\partial \boldsymbol{\theta}}.
\end{split}
\end{equation}

Once $\sans{K}_{y}^{-1}(\bs{y} - \bs{\Phi}\hat{\bs{\beta}})$ is precomputed (after learning the kernel hyperparameters, $\bs{\nu}$) and stored, the first and second derivatives can be computed very quickly. In the case of finite difference methods, if a poor finite step size is specified, numerical derivatives can become unstable. This is not the case in this framework. In Figure \ref{fig:gradients}, we show the first derivatives with respect to the input cosmological parameters, $\bs{\theta}=(\Omega_{\tm{cdm}}h^{2},\,\Omega_{\tm{b}}h^{2},\,\tm{ln}(10^{10}A_{s}),\,n_{s},\,h)$. The first derivatives with CLASS (in red) are calculated using finite central difference method. 

%% file: Content/weaklensing.tex
\section{Weak Lensing Power Spectra}
\label{sec:wl}
A crucial application of the 3D matter power spectrum is in a weak lensing analysis, where the calculation of the different power spectra types is required. In the absence of systematics, most of the cosmological information lies in the curl-free (E-) component of the shear field. The Limber approximation \citep{1953ApJ...117..134L, 2008PhRvD..78l3506L} is typically assumed and under the assumption of no systematics, the E-mode lensing power spectrum is equal to the convergence power spectrum and is given by:

\begin{equation}
\label{eq:ee_ps}
C_{\ell,\,ij}^{\tm{EE}}=\int_{0}^{\chi_{H}}\tm{d}\chi\,\dfrac{w_{i}(\chi)w_{j}(\chi)}{\chi^{2}}\,P_{\delta}^{\tm{bary}}(k,\chi).
\end{equation}

\noindent and 

\begin{equation}
w_{i}(\chi)=A\chi(1+z)\int_{\chi}^{\chi_{\tm{H}}}\tm{d}\chi'\,n_{i}(\chi)\left(\dfrac{\chi'-\chi}{\chi'}\right)
\end{equation}

\noindent where $A=3H_{0}^{2}\Omega_{\tm{m}}/(2c^{2})$. $\chi$ is the comoving radial distance, $\chi_{\tm{H}}$ is the comoving distance to the horizon, $H_{0}$ is the present day Hubble constant and $\Omega_{\tm{m}}$ is the matter density parameter. $w_{i}$ is the weight function which depends on the lensing kernel. The weight function is a measure of the lensing efficiency for tomographic bin $i$. Moreover, the redshift distribution, $n_{i}(z)$, as a function of the redshift, is related to the comoving distance via a Jacobian term, that is, $n(z)\;\tm{d}z=n(\chi)\;\tm{d}\chi$ and it is also normalised as a probability distribution, that is, $\int n(z)\;\tm{d}z = 1$.

\subsection{Intrinsic Alignment Power Spectra}
An important theoretical astrophysical challenge for weak lensing is intrinsic alignment (IA). It gives rise to preferential and coherent orientation of galaxy shapes, not because of lensing alone but due to other physical effects. Although not very well understood, it is believed to arise by two main mechanisms, namely the interference (GI) and intrinsic alignment (II) effects, such that the total signal is in fact a biased tracer of the true underlying signal, $C_{\ell,ij}^{\tm{EE}}$, that is,

\begin{equation}
\label{eq:cl_tot}
C_{\ell,ij}^{\tm{tot}}=C_{\ell,ij}^{\tm{EE}}+A_{\tm{IA}}^{2}C_{\ell,ij}^{\tm{II}}-A_{\tm{IA}}C_{\ell,ij}^{\tm{GI}}
\end{equation}

\noindent where $A_{\tm{IA}}$ is a free amplitude parameter, which allows for the flexibility of varying the strength of the power, arising due to the intrinsic alignment effect. In particular, the II term arises as a result of alignment of a galaxy in its local environment whereas the GI term is due to the correlation between the ellipticities of the foreground galaxies and the shear of the background galaxies. Note that, the II term contributes positively towards the total lensing signal whereas the GI subtracts from the signal. The II power spectrum is given by

\begin{equation}
\label{eq:ii_ps}
C_{\ell,ij}^{\tm{II}}=\int_{0}^{\chi_{\tm{H}}}\tm{d}\chi\;\dfrac{n_{i}(\chi)\,n_{j}(\chi)}{\chi^{2}}\,P_{\delta}^{\tm{bary}}(k,\chi)\;F^{2}(\chi)
\end{equation}

\noindent and the GI power spectrum is

\begin{equation}
\label{eq:gi_ps}
C_{\ell,ij}^{\tm{GI}}=\int_{0}^{\chi_{\tm{H}}}\tm{d}\chi\,\dfrac{w_{i}(\chi)n_{j}(\chi)+w_{j}(\chi)n_{i}(\chi)}{\chi^{2}}\,P_{\delta}^{\tm{bary}}(k,\chi)\,F(\chi),
\end{equation}

\noindent where $F(\chi)=C_{1}\rho_{\tm{crit}}\Omega_{\tm{m}}/D(\chi)$. $D(\chi)$ is the linear growth factor normalised to unity today, $C_{1}=5\times10^{-14}\;h^{-2}\tm{M}_{\odot}^{-1}\tm{Mpc}^{3}$ and $\rho_{\tm{crit}}$ is the critical density of the Universe today. As seen from Equations \ref{eq:ee_ps}, \ref{eq:ii_ps} and \ref{eq:gi_ps}, they all involve an integration of the form

\begin{equation}
\label{eq:wl_general}
C_{\ell}=\int_{0}^{\chi_{\tm{H}}}\;g(\chi)\,P_{\delta}^{\tm{bary}}(k,\chi)\;\tm{d}\chi.
\end{equation}

Hence, an emulator for $P_{\delta}(k,z)$ will enable us to numerically compute all the weak lensing power spectra in a fast way. This will be useful in future weak lensing surveys where we will require many power spectra calculation as a result of the large number of auto- and cross- tomographic bins. For example, in the recent KiDS-1000 analysis \citep{2021A&A...645A.104A}, five tomographic bins were employed, resulting in 15 (multiplied by 3 if we are including intrinsic alignment power spectra) power spectra calculations. In future surveys, it is expected that the number of redshift bins will be of the order 10, thus requiring at least 55 power spectra calculations for each power spectrum type (EE, GI and II).

\subsection{Redshift Distribution}
An important quantity for calculating the weak lensing power spectra is the redshift distribution. For an in-depth cosmological data analysis such as the Kilo Degree Survey (KiDS), it is crucial to calibrate the photometric redshift to obtain robust model predictions. For more advanced techniques for estimating the $n(z)$ from photometric redshifts, we refer the reader to techniques such as weighted direct calibration, DIR \citep{2008MNRAS.390..118L, 2016PhRvD..94d2005B}, calibration with cross-correlation, CC \citep{2008ApJ...684...88N} and recalibration of photometric $P(z)$, BOR by \cite{2010MNRAS.406..881B}. Recently \cite{2016MNRAS.460.4258L} developed a hierarchical Bayesian inference method to infer redshift distributions from photometric redshifts.

\begin{figure}[H]
\noindent \begin{centering}
\includegraphics[width=0.45\textwidth]{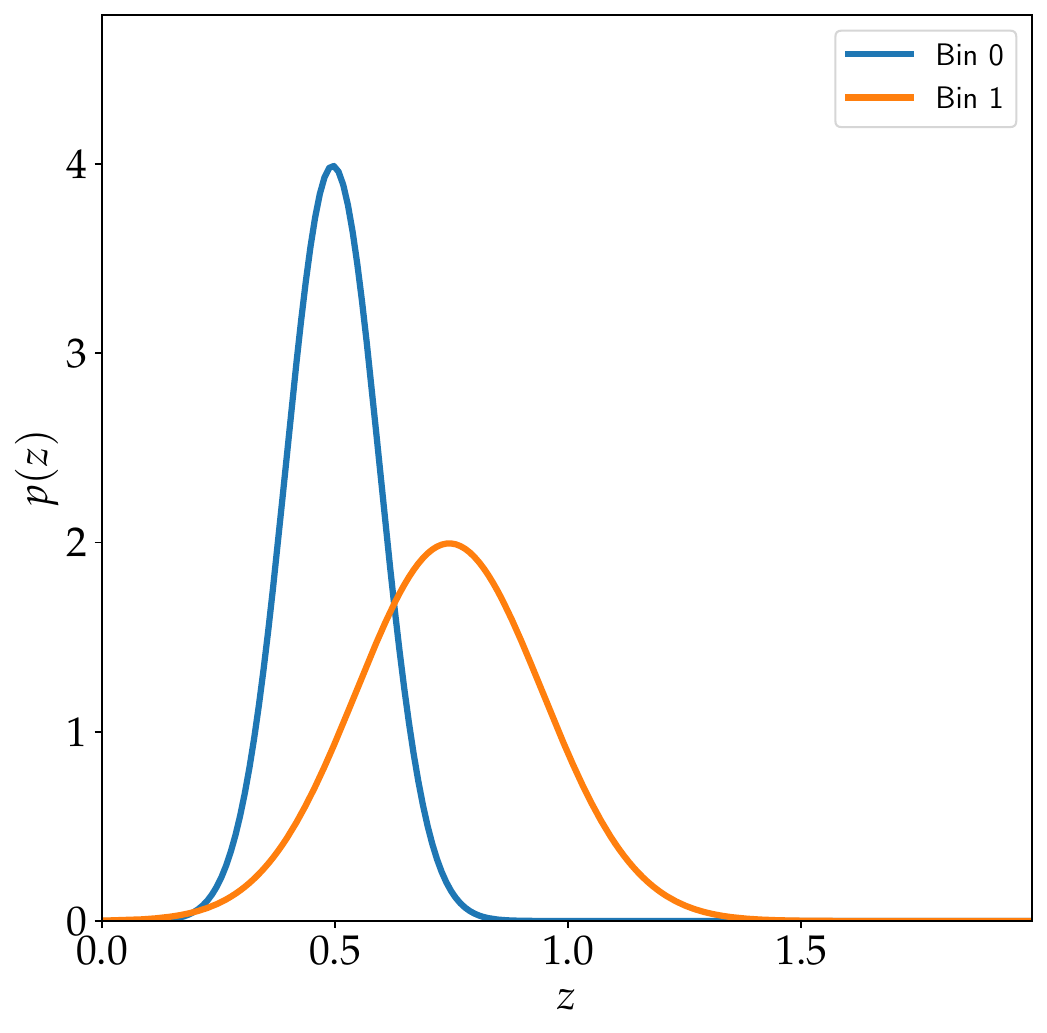}
\par\end{centering}
\caption{\label{fig:nz_dist}To illustrate the calculation of the weak lensing power spectra, we use two analytic redshift distributions centred at redshift 0.50 and 0.75 respectively. The $n(z)$ distribution assumed here is a normal distribution and is given by Equation \ref{eq:nz_gaussian}. The standard deviations for each normal distribution are set to 0.1 and 0.2 respectively.}   
\end{figure}

In this work, we use a toy Gaussian distribution to illustrate how we can use the 3D matter power spectrum, $P_{\delta}(k,z)$ in conjunction with the $n(z)$ distribution to calculate the different weak lensing power spectra. Note that one can just replace this toy $n(z)$ distribution example by any redshift distribution as calculated by any one of the techniques mentioned above.

Different $n(z)$ distributions are available as part of the software. The first 2 distributions are:

\begin{equation}
\label{eq:nz_model_1}
n(z) = B\,z^{2}\tm{exp}\left(-\dfrac{z}{z_{0}}\right)
\end{equation}

\noindent and 

\begin{equation}
\label{eq:nz_model_2}
n(z) = B\,\alpha z\tm{exp}\left[-\left(\dfrac{z}{z_{0}}\right)^{\beta}\right].
\end{equation}

\noindent For a \textit{Euclid}-like survey, $z_{0}\sim 0.7$, $\alpha=2$ and $\beta = 1.5$ \citep{2015MNRAS.449.1146L}. The third distribution implemented is just a Gaussian distribution with mean $z_{0}$ and standard deviation, $\sigma$

\begin{equation}
\label{eq:nz_gaussian}
n(z) = B\,\tm{exp}\left[-\dfrac{1}{2}\left(\dfrac{z-z_{0}}{\sigma}\right)^{2}\right]
\end{equation}

\noindent where $B$ is a normalisation factor such that $\int n(z)\;\tm{d}z=1$ in all cases above. As shown in Figure \ref{fig:nz_dist}, we employ two redshift distributions, where the mean and standard deviation for the first distribution are 0.50 and 0.10 respectively and for the second distribution (in orange), the mean and standard deviation are set to 0.75 and 0.20 respectively.

%% file: Content/software.tex
\section{Software}
\label{sec:software}

In this section, we briefly elaborate on how the code is set up and the different functionalities one can exploit. Note that any default values mentioned below can be adjusted according to the user's preferences. The default values of the minimum and maximum redshifts are set to 0 and 4.66 respectively and as discussed in \S\ref{sec:procedures}, we also assume 20 redshifts spaced equally in the linear scale. For the wavenumbers in units of $h\,\tm{Mpc}^{-1}$, the minimum is set to $5\times 10^{-4}$ and the maximum to 50, with 40 wavenumbers equally spaced in logarithmic scale. A fixed neutrino mass of 0.06~eV is assumed but this can also be fixed at some other value or it can also be included as part of the emulation strategy. The code supports either choice.

\begin{figure}[H]
\noindent \begin{centering}
\includegraphics[width=0.45\textwidth]{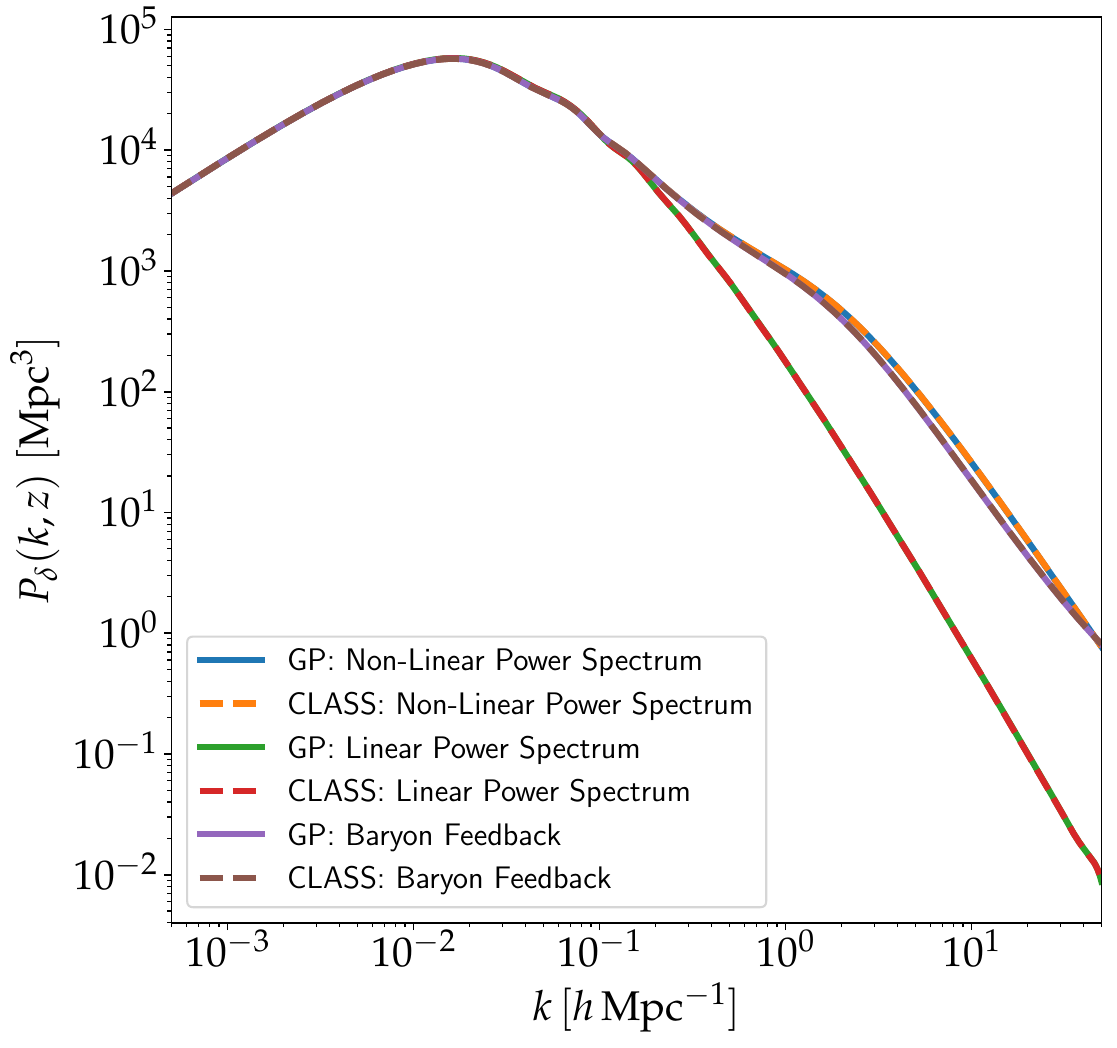}
\par\end{centering}
\caption{\label{fig:pk_nl_gp_class}The linear power spectrum at a fixed redshift, $z_{0}$, the 3D non-linear matter power spectrum, $P_{\delta}(k,z)$ and the 3D non-linear matter power spectrum with baryon feedback, $P_{\delta}^{\tm{bary}}(k,z)$ can be calculated with our emulating scheme. The solid curves correspond to predictions from the model while the broken curves show the accurate functions as calculated with CLASS.}   
\end{figure}

The next step involves generating the training points. We generate 1000 LH design points using the \texttt{maximinLHS} function and we calculate and record the three quantities, the growth factor, $D(z)$, the non-linear function, $q(k,z)$ and the linear matter power spectrum, $P_{\tm{lin}}(k,z_{0})$. At a very small value of $k$, which we refer to as $k_{\tm{min}}$, $q=0$. The non-linear matter power spectrum is only relevant for some range of $k_{nl}$ and $k_{nl}>k_{\tm{min}}$. Hence, the growth factor is just

\begin{equation}
    D(z)=\dfrac{P_{\tm{lin}}(k_{\tm{min}},z)}{P_{\tm{lin}}(k_{\tm{min}}, z_{0})}
\end{equation}

\begin{figure}[H]
\noindent \begin{centering}
\includegraphics[width=0.45\textwidth]{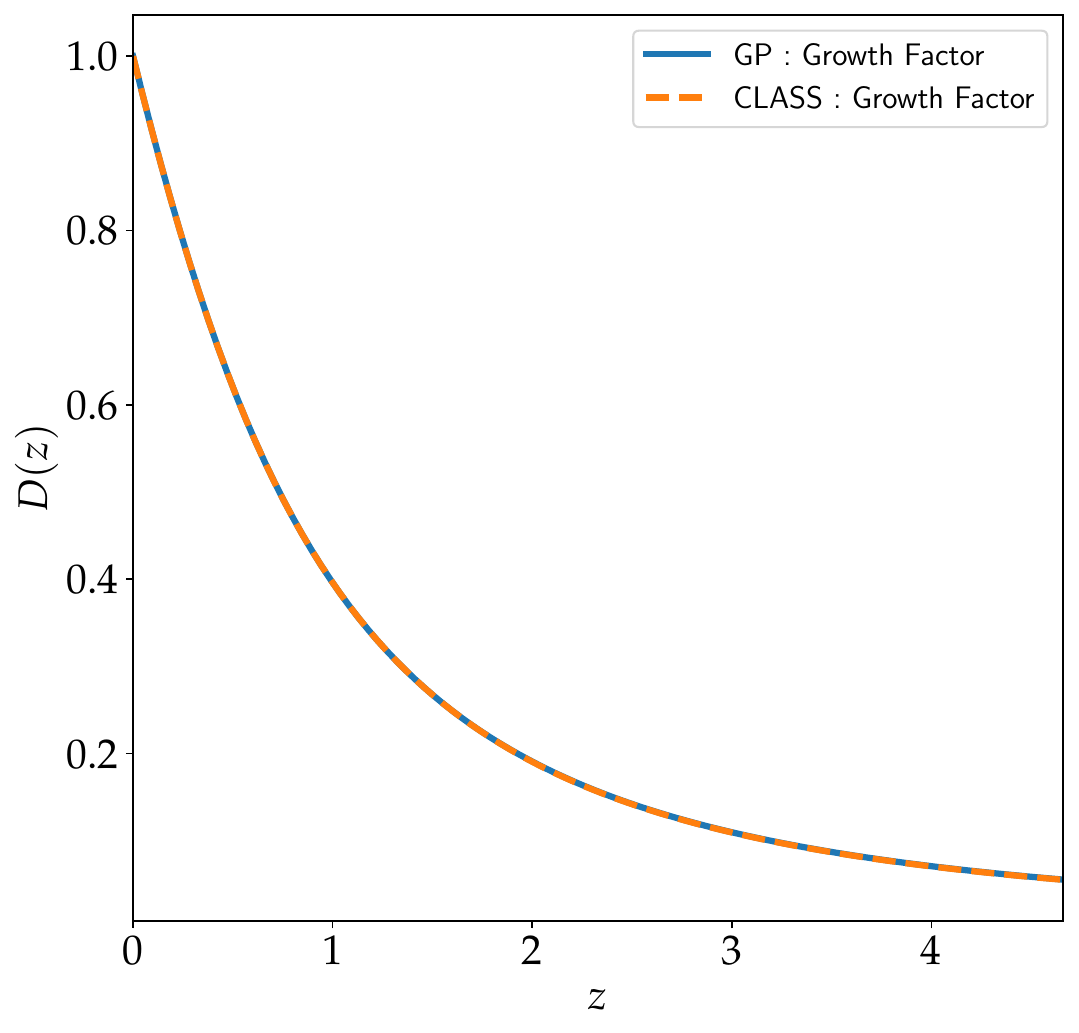}
\par\end{centering}
\caption{\label{fig:gf_gp_class}The growth factor, $D(z)$ as predicted by the surrogate model (in blue) at a test point in parameter space. The accurate function is also calculated using CLASS and is shown in orange. Recall that the emulator is constructed for $z\in [0.0, 4.66]$, aligned with current weak lensing surveys.}   
\end{figure}

\noindent Throughout our analysis, we use $z_{0}=0$. In some regions of the parameter space, we also found that the $q(k,z)$ were noisy and this can be alleviated by increasing the parameter \texttt{P\_k\_max\_h/Mpc} when running CLASS. If a small value is assumed, the interpolation in the high-dimensional space will not be robust. We set this value to 5000 to ensure the $q(k,z)$ function remains smooth as a function of the inputs. However, this procedure leads to CLASS being slower. It takes $\sim 30$ seconds on average to do 1 forward simulation. For example, in our application, it took 520 minutes to generate the targets $(D,\,q,\,P_{\tm{lin}})$ for 1000 input cosmologies. We have also found that CLASS occasionally fails to compute the power spectrum and this is resolved as follows. We allocate a time frame (60 seconds in this work) for CLASS to attempt to calculate the power spectrum and if it fails, a small perturbation is added to the input training point parameters and we re-run CLASS, until the power spectrum is successfully calculated. In the failing cases, the maximum number of attempts is only 3. Moreover, the code currently supports polynomial functions of order 1 and 2, that is, the set of basis functions for an order 2 polynomial is $[1,\,\bs{\theta},\,\bs{\theta}^{2}]$. For example, \cite{2011ApJ...728..137S} implemented a first and second order polynomial function to design an emulator for the CMB while \cite{2007ApJ...654....2F} used a fourth order polynomial function. In this case, recall that we are also marginalising over the residuals analytically by using the kernel function. Training the emulator, that is, learning the kernel hyperparameters, for the different targets, took around 340 minutes. All experiments were conducted on an Intel Core i7-9700 CPU desktop computer.

Note that we do not compute the emulator uncertainty for various reasons. Simulators such as CLASS are deterministic input-output models, that is, running the simulator again at the same input values will give the same outputs and, as argued by \cite{doi:10.1198/TECH.2009.08019}, the error returned by the GP is unreliable \citep{2020MNRAS.497.2213M}. 

Moreover, the emulator uncertainty changes as a function of the number of training points and so do the accuracy and precision of the predicted mean function from the emulator. However, in a small-data regime --- for example, band powers for current weak lensing surveys --- the emulator uncertainty might have significant undesirable effects on the inference of the cosmological parameters. On a more technical note, storing and calculating the emulator uncertainty is a demanding process, both with $\mc{O}(N^{2})$ computational cost respectively, where $N$ is the number of training points.

Once all these processes (generating the training points and training the emulators) are completed, the emulator is very fast when we compute the 3D matter power spectrum. It takes around 0.1 seconds to do so compared to the average value of 30 seconds by CLASS. Note that the gradient calculation with the emulator is even more efficient compared to finite difference methods, where CLASS would need to be called 10 times for a 5D problem (assuming a central difference method). For an in-depth documentation on the code structure and technical details, we refer the reader to \S\ref{sec:code_availability}, where we provide the links to the code and documentation.

%% file: Content/results.tex
\section{Results}
\label{sec:results}

In this section, we highlight the main results, starting from the calculation of the 3D matter power spectrum to the calculation of the different weak lensing power spectra. 

In Figure \ref{fig:gradients}, we show the gradient along at a fixed cosmological parameter (test point) and fixed redshift, $z=0$. The red curve corresponds to the gradients as calculated by CLASS using central difference method and the blue curves show the gradients output from the emulator. In particular, this gradient is strictly a 3D quantity, as a function of the wavenumber $k$, redshift, $z$ and the cosmological parameters $\bs{\theta}$. In other words, the gradient calculation from the emulator will be a tensor of size $(N_{k},\,N_{z},\,N_{p})$, where $N_{k}$ is the number of wavenumbers for $k\in[5\times 10^{-4},\,50]$, $N_{z}$ is the number of redshifts for $z\in[0.0,\,4.66]$ and $N_{p}$ is the number of parameters considered. In this case, $N_{p}=5$ and the default values for a finer grid in $k$ and $z$ are $N_{k}=1000$ and $N_{z}=100$.

In Figure \ref{fig:gf_gp_class}, we show the growth factor, $D(z)$ calculated using CLASS (in orange) and the emulator (in blue), while in Figure \ref{fig:pk_nl_gp_class}, we show three important quantities. First, since we are emulating the 3 different components of the non-linear matter power spectrum, we are able to compute the linear matter power spectrum at a test point, at the reference redshift, $z_{0}=0$. Note that the one calculated by CLASS and the one by the emulator agree quite well. Similarly, we can also calculate the 3D non-linear matter power spectrum and in Figure \ref{fig:pk_nl_gp_class}, in orange and blue, we have the power spectrum at a fixed redshift, excluding baryon feedback, calculated using CLASS and the emulator respectively. The same is repeated for the curves in purple and brown, but in this case including baryon feedback. As discussed in \S\ref{sec:model}, we can also see the effect of baryon feedback which alters the power spectrum at large $k$.

\begin{figure}[H]
\noindent \begin{centering}
\includegraphics[width=0.45\textwidth]{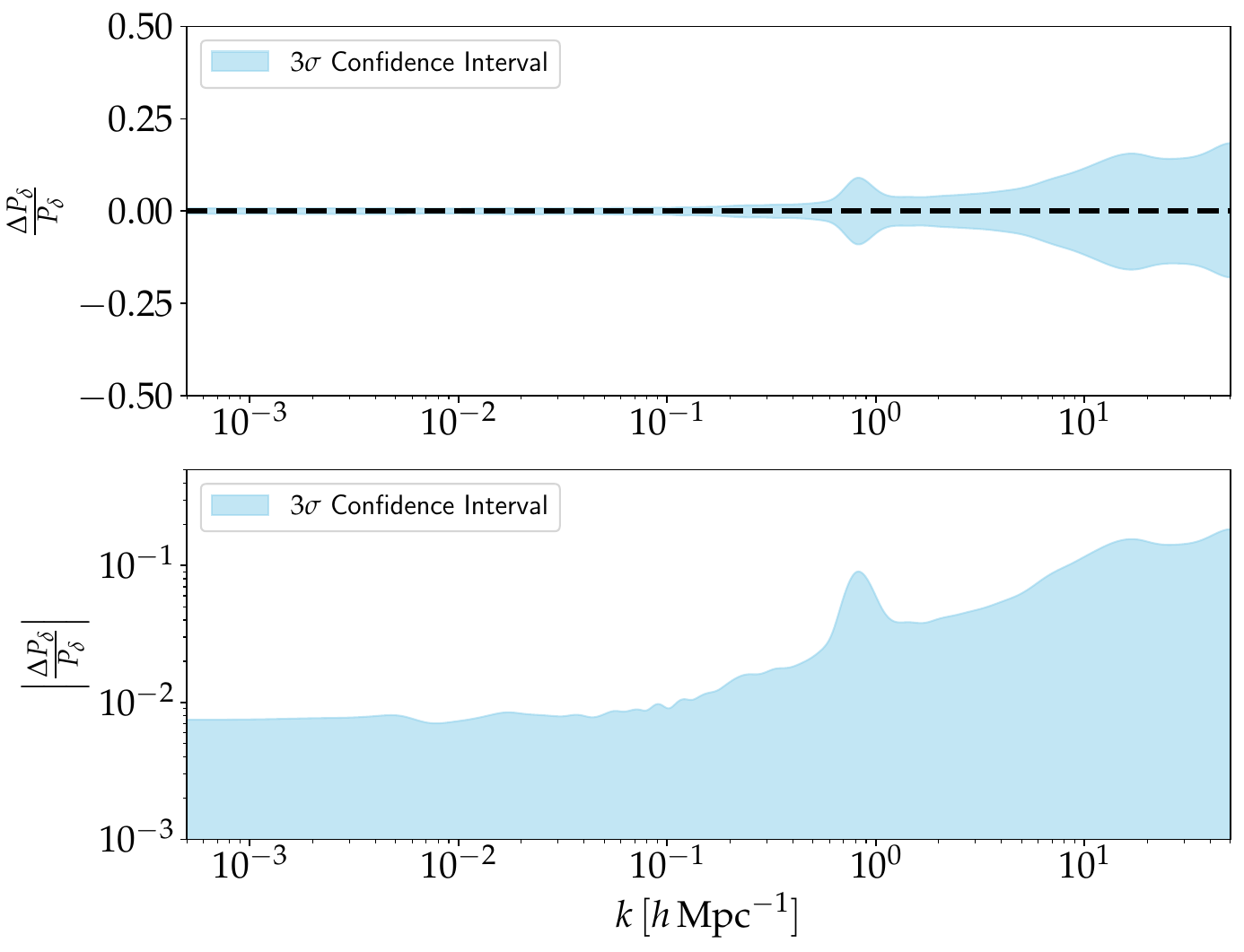}
\par\end{centering}
\caption{\label{fig:delta_p}To investigate the performance of the emulator, we draw an independent set of cosmological parameters, randomly from the prior and we calculate the fractional error between the predicted ones with the GP model and CLASS. The mean of $\nicefrac{\Delta P_{\delta}}{P_{\delta}}$ is shown by the broken horizontal line and the $3\sigma$ confidence interval, derived from the standard deviations of all experiments, is shown in pale blue. For an accurate emulator, it is expected that the mean is centred on 0 and this demonstrates the robustness of this method. Note that in this procedure, one can also specify the number of desired power spectra for $z\in[0.0, 4.66]$. For example, for $p$ cosmological parameters and $n$ redshifts, we have $np$ power spectra outputs. In the bottom panel, we show the absolute error on a logarithmic scale.}  
\end{figure}

\begin{figure*}
    \centering
    \subfloat[The EE power spectrum, $C_{\ell,ij}^{\tm{EE}}$]{{\includegraphics[width=0.30\textwidth]{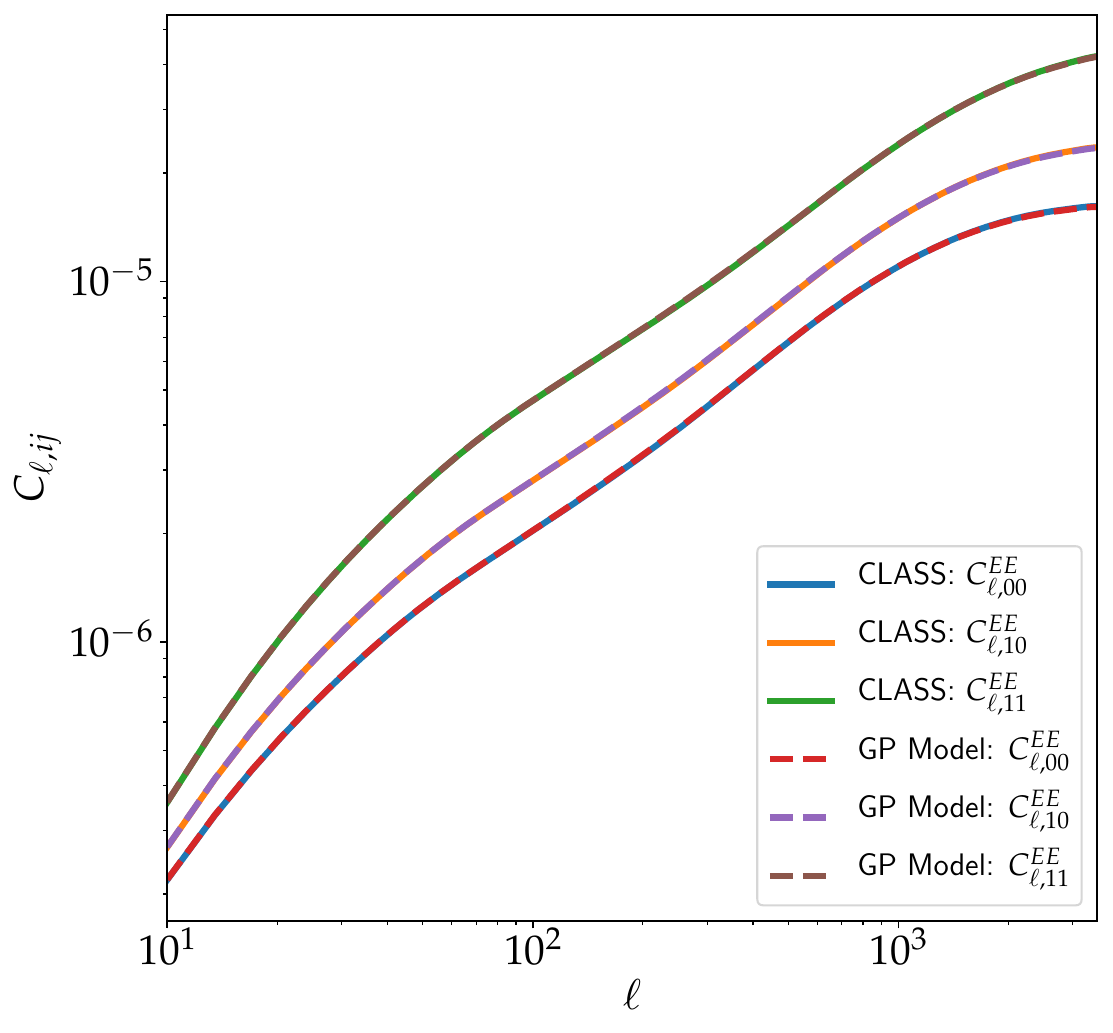}}}
    \qquad
    \subfloat[The II power spectrum, $C_{\ell,ij}^{\tm{II}}$]{{\includegraphics[width=0.30\textwidth]{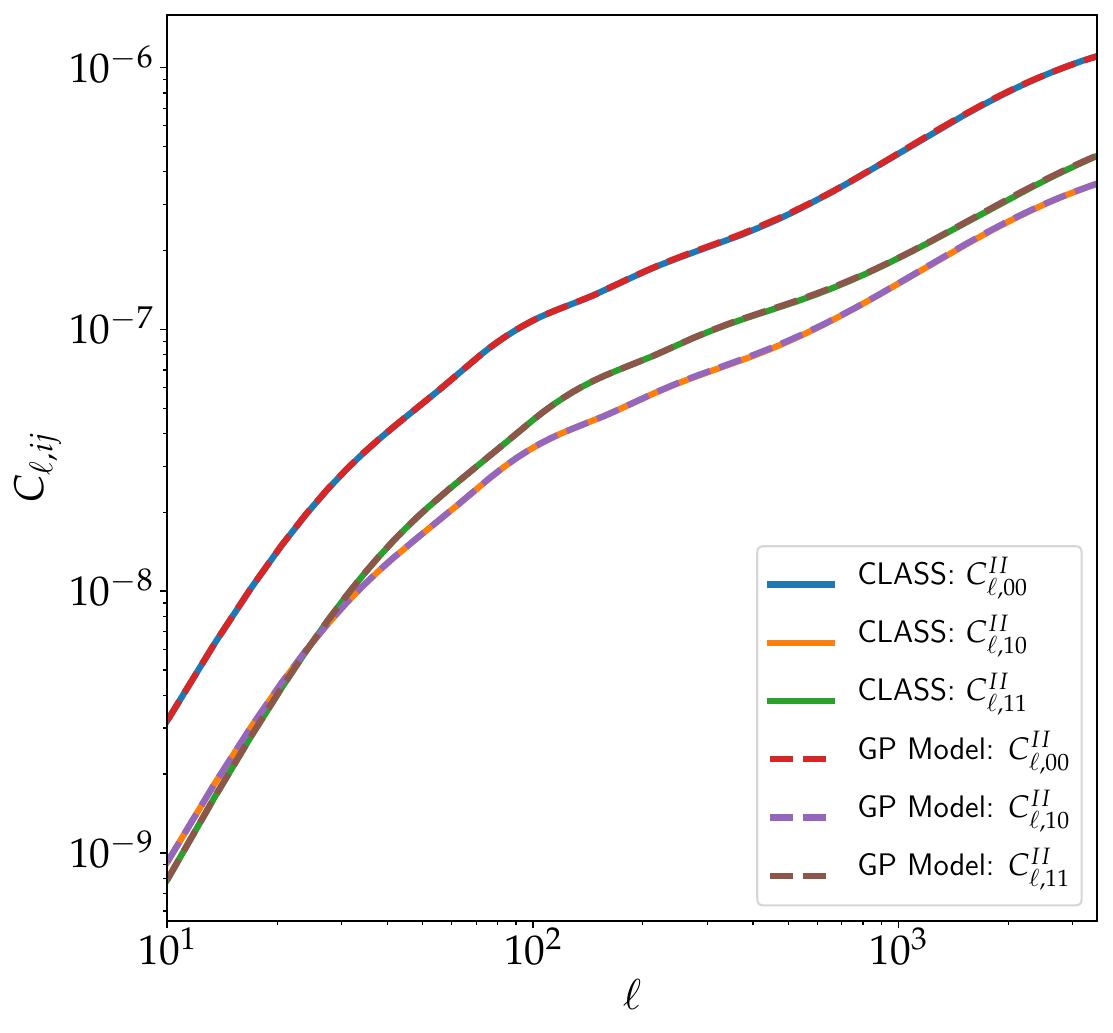}}} 
    \qquad
    \subfloat[The GI power spectrum, $C_{\ell,ij}^{\tm{GI}}$]{{\includegraphics[width=0.30\textwidth]{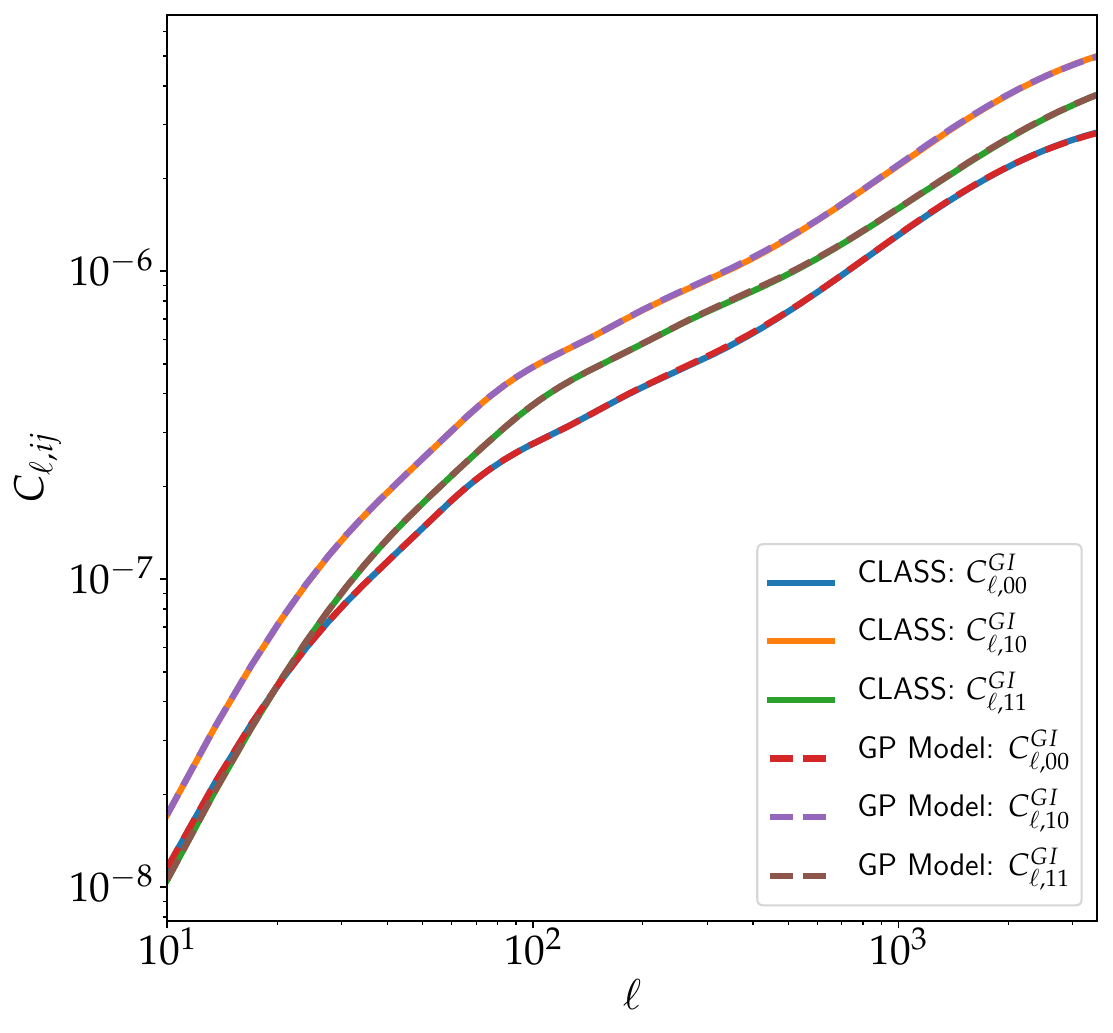}}} 
    \caption{\label{fig:ee_ii_gi_gp_class}The left, centre and right panels show the different weak lensing power spectra as calculated by the emulator (broken curves) and the accurate model, CLASS, shown by the solid curves. The different power spectra within each panel correspond to the auto- and cross- power spectra, due to the 2 tomographic redshift distribution in Figure \ref{fig:nz_dist}, hence leading to 00, 10, and 11 power spectra. These power spectra are then added, via the intrinsic alignment parameter, $A_{\tm{IA}}$ to construct a final model, $C_{\ell,ij}^{\tm{tot}}$ in a weak lensing analysis. See Equation \ref{eq:cl_tot}.}
\end{figure*}

Various techniques have been proposed by \cite{doi:10.1198/TECH.2009.08019} to assess the performance of an emulator. These diagnostics are generally based on the comparisons between the emulator and simulator runs for new test points in input parameter space. These test points should cover the input parameter space over which the training points were previously generated. In this application, we randomly choose 100 independent test points from the prior range and evaluate the simulator and the emulator at these points. Since, we are emulating the 3D matter power spectrum, we can also generate it on a finer grid, unlike the previous setup where we used 40 wavenumbers and 20 redshifts. Hence, we generate all the power spectra for 1000 wavenumbers, equally spaced in logarithmic scale, $k\in [5\times 10^{-4},\,50]$ and 100 redshifts, $z\in[0.0,\,4.66]$, equally spaced in linear scale. For the 100 test points, this gives us a set of $10^{4}$ power spectra. We define the fractional uncertainty as

\begin{equation}
\dfrac{\Delta P_{\delta}}{P_{\delta}}=\dfrac{P_{\delta}^{\tm{emu}} - P_{\delta}}{P_{\delta}}
\end{equation}

\noindent and given the set of power spectra we have generated, we compute the mean and variance of $\nicefrac{\Delta P_{\delta}}{P_{\delta}}$. For a robust emulator, the mean should be centred on zero and indeed, as seen, from Figure \ref{fig:delta_p}, the mean is centred on 0. The variance, depicted by the $3\sigma$ confidence interval in pale blue, is also quite small. 

\begin{figure}[H]
\noindent \begin{centering}
\includegraphics[width=0.45\textwidth]{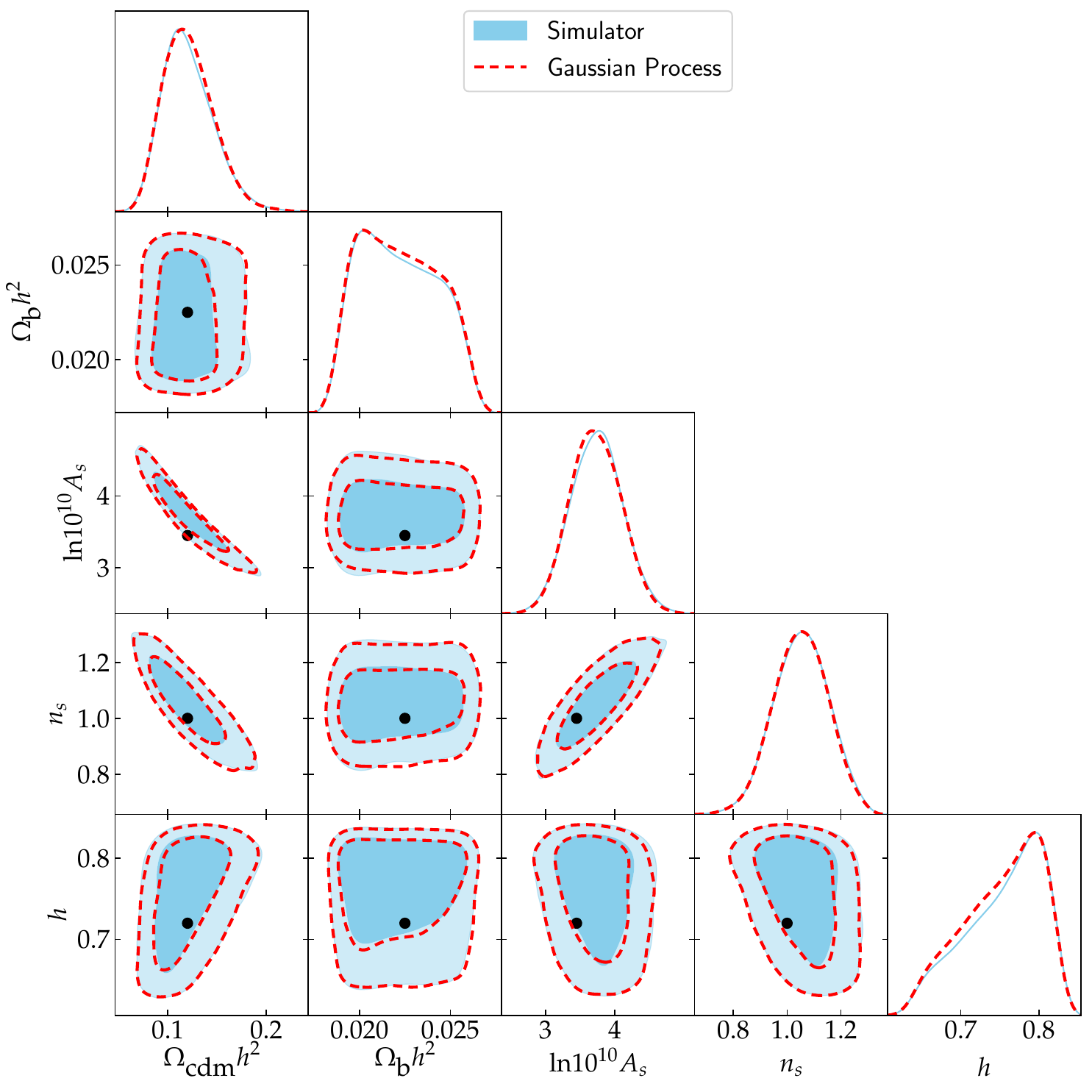}
\par\end{centering}

\caption{\label{fig:inference_example}Marginalised posterior distributions of the five cosmological parameters. The blue colour refers to the posterior distribution of the parameters as inferred using CLASS and the broken red contours refer to the posterior distribution when using the emulator developed in this work. The black dots correspond to the fiducial point in parameter space where the data have been generated.}  
\end{figure}

In Figure \ref{fig:ee_ii_gi_gp_class}, we show the different types of weak lensing power spectra calculated using CLASS and the emulator. The left, middle and right panel show the auto- and cross- EE, II and II power spectra due to the two tomographic bins, shown in Figure \ref{fig:nz_dist}. In the three panels, the blue, orange and green curves correspond to the auto- and cross- power spectra, $C_{\ell,00}$, $C_{\ell,10}$ and $C_{\ell,11}$ as computed by CLASS. Similarly, the red, purple and brown broken curves are the power spectra generated by the emulator. The power spectra are in agreement when comparing CLASS and the emulator. Note that, in a typical weak lensing analysis, the three different types of power spectra (EE, GI and II) are combined together via the intrinsic alignment parameter, $A_{\tm{IA}}$ (see Equation \ref{eq:cl_tot}). 

We also test the emulator on simulated weak-lensing bandpowers. We assume measurements over $10\leq\ell\leq 1500$ and 5 tomographic slices with Gaussian $n(z)$ as in Equation \ref{eq:nz_gaussian}, centred on redshifts [0.5, 1.0, 1.5, 2.0, 2.5] and each having a standard deviation of 0.075. Ten bandpowers, equally spaced in logarithmic scale, are used and this gives us a set of 150 data points. Moreover, we simulate and then assume in the likelihood  independent Gaussian errors with, for simplicity, $\sigma=0.5\hat{\mc{B}}_{\ell}$, where $\hat{\mc{B}}_{\ell}$ is the bandpower evaluated at the fiducial set of cosmological parameters. For this particular case, we have set $A_{\tm{IA}}=0$ but one can trivially include this factor and marginalise over it in the sampling process. The fiducial point $\bs{\theta}_{\tm{fid}}=[0.12, 0.0225, 3.45, 1.0, 0.72]$ is used to generate the data and is shown by the black dots in Figure \ref{fig:inference_example}. We use a Gaussian likelihood and uniform priors on all cosmological parameters, similar to the range of the inputs of the emulator. Figure \ref{fig:inference_example} shows the results obtained when sampling the cosmological parameters on this toy data set. The red contours correspond to the result using the emulator while the pale blue colour refers to the posterior distributions using CLASS. We run three separate MCMC chains, each with 150\,000 MCMC samples, two with the emulator and one with CLASS. On each of the three resulting pairs of runs, we compute the Gelman-Rubin convergence parameter, $\hat{R}$ \citep{1992StaSc...7..457G}. The worst $\hat{R}$ value is 1.002, consistent with all three chains being drawn from the same distribution, and corroborating the agreement shown in Figure \ref{fig:inference_example}.  The emulator developed in this work is thus able to robustly recover the posterior distributions of all the cosmological parameters, compared to the accurate solver, CLASS.

%% file: Content/conclusions.tex
\section{Conclusions}
\label{sec:conclusions}

In this paper, we have proposed an emulator for the 3D matter power spectrum as calculated by CLASS across a wide range of cosmological parameters (see Table \ref{tab:prior_range}). This detailed methodology presented in this work entails a multifaceted view of the 3D power spectrum, which is an essential quantity in a weak lensing analysis. In particular, we have successfully demonstrated that as part of this routine, we can compute the linear matter power spectrum at a reference redshift $z_{0}$, the non-linear 3D matter power spectrum with and without the baryon feedback model described in \S\ref{sec:model}, gradients of the 3D matter power spectrum with respect to the input parameters and the different auto- and cross- weak lensing power spectra (EE, GI and II) derived from $P_{\delta}^{\tm{bary}}(k,z)$ and the given tomographic redshift distributions, $n_{i}(z)$. Note that the gradients of the weak lensing power spectra are also straightforward to calculate using the distributive property of gradients (see Eq.~\ref{eq:wl_general} for a general form for the different weak lensing power spectra). Note that only $P_{\delta}^{\tm{bary}}(k,z)$ is a function of the cosmological parameters.

The default emulator is built using 1000 training points only and because the mean of the surrogate model is just a linear predictor, the mean function is very quick to compute. In the same spirit, the first and second derivatives involve only element-wise matrix multiplication, and are therefore quick to compute. In the test cases, a full 3D matter power spectrum calculation takes 0.1 seconds compared to an average value of 30 seconds when CLASS is used. While the goal remains to have an emulating method which is faster than the computer model, it is also worth pointing out that it also quite accurate, following the diagnostics we have performed in this work, see Figure \ref{fig:delta_p} as an example. The emulator can be made more accurate and precise as we add more and more training points, but this comes at an expense of $\mc{O}(N^{3})$ cost at each optimisation step during the training phase. Fortunately, in this work, 1000 training points suffice to yield promising and robust power spectra.

Building an emulator for the 3D power spectrum is deemed to be a challenging task \citep{2020PhRvD.102f3504K}, the main difficulties arising due to the fact that GP models cannot easily handle large datasets ($\sim 10^4$ training points) and it is not trivial to work with vector-valued functions, for example, $P_{\delta}(k,z;\bs{\theta})$ as in this work. Also, techniques such as multi-outputs GP result in large matrices, hence a major computational challenge. Fortunately, the method presented in this work, along with the projection method explained in \S\ref{sec:training_points}, provides a simple and straightforward path towards building emulators. 

Moreover, current weak lensing data do not constrain the cosmological parameters to a high precision, hence motivating us to distribute 1000 training points across a large parameter space, according to the current prior distributions (hypercube) used in the literature. In future weak lensing surveys, with improved precision on the parameters, one can choose to use, for example, a multi-dimensional Gaussian prior (hypersphere) which will certainly have a much smaller volume compared to the hypercube used in this work. If we stick with 1000 training points, this will lead to very precise power spectra, or we can also opt to distribute fewer than 1000 training point across the parameter space. Fewer training points also imply that training the emulator will be faster.

The different aspects of the emulation scheme proposed in this work can easily pave their way into different cosmological data analysis problems. A nice example is an analysis combining the MOPED data compression algorithm \citep{2000MNRAS.317..965H}, the emulated 3D matter power spectrum and the $n(z)$ uncertainty in a weak lensing analysis. Moreover, if we want to use a more sophisticated sampler such as Hamiltonian Monte Carlo (HMC), one can leverage the gradients from the emulator to derive an expression for the gradient of the negative log-likelihood (the potential energy function in an HMC scheme) with respect to the input cosmological parameters, under the assumption that such an analytic derivation is possible. Furthermore, the second derivatives can be used in a Fisher Matrix analysis, or the first and second derivatives can be be used together in an approximate inference scheme based on Taylor expansion techniques, see for example, the recent work by \cite{2019MNRAS.490.4237L}. In addition, similar concepts behind this work can be extended to build emulators for $P_{\delta}(k,z)$ from N-body simulations.

%% file: Content/acknowledgement.tex
\section*{Acknowledgement}
\label{sec:acknowledgement}
We would like to thank Zafiirah Hosenie for insightful discussions on the Machine Learning model building and training process. We also thank colleagues Boris Leistedt and George Kyriacou for providing feedback and discussing this work during a presentation. A.M is supported financially by the Imperial College President's scholarship.

%% file: Content/code.tex
\section*{Data Availability}
\label{sec:code_availability}
The code, written in Python, is publicly available on Github at \href{https://github.com/Harry45/emuPK}{https://github.com/Harry45/emuPK} and the documentation is maintained at \href{https://emupk.readthedocs.io/}{https://emupk.readthedocs.io/}. The trained surrogate models are also made available as part of this distribution. One can also follow the instructions in the documentation to train her own emulator based on the desired configurations.